\documentclass{jfm}
\pdfoutput=1
\usepackage{amsmath,amssymb,graphicx,bm}  
\usepackage{physics,mathtools,scalerel}
\usepackage{tikz}
\usetikzlibrary{shapes,calc}

\newcommand{\overbar}[1]{\mkern 1.5mu\overline{\mkern -1.5mu#1 \mkern -1.5mu}\mkern 1.5mu}
\newcommand{\ieps}{\frac{1}{\epsilon}}
\newcommand{\Ret}{\Rey_{\tau}}
\newcommand{\inv}{^{\raisebox{.2ex}{$\scriptscriptstyle-1$}}}
\newcommand{\mnU}{\overbar{U}}

\newcommand{\kpyy}{\sqrt{\smash{K_y^+}\vphantom{K_x^+}}}
\newcommand{\kpxx}{\sqrt{\smash{K_x^+}\vphantom{K_x^+}}}
\newcommand{\kpzz}{\sqrt{\smash{K_z^+}\vphantom{K_x^+}}}

\newcommand{\llx}{\lambda_x^+}
\newcommand{\llz}{\lambda_z^+}
\newcommand{\kxx}{\kappa_x}
\newcommand{\kzz}{\kappa_z}
\newcommand{\bkv}{\bm{\kappa}}
\newcommand{\bfu}{\boldsymbol{u}}              
\newcommand{\bfU}{\overbar{\boldsymbol{U}}}    
\newcommand{\tnabla}{\tilde{\grad}}
\newcommand{\bfK}{\mathsfbi{K}}
\newcommand{\bbfu}{\langle\boldsymbol{u}\rangle}  
\newcommand{\diagK}{\text{diag}({K}_{x},{K}_{y},{K}_{z})}
\newcommand{\WmatU}{\mathsfbi{W}_{\boldsymbol{u}}}
\newcommand{\WmatF}{\mathsfbi{W}_{\boldsymbol{f}}}

\renewcommand{\epsilon}{\varepsilon}

\begin{document}
\newcommand{\bcircle}{\raisebox{-0.5pt}{\tikz{
			\draw[-,black,line width = 0.9pt](-2mm,0)--(2mm,0);
			\draw[line width=1pt,fill=white] (0,0) circle (3pt);}}\phantom{,}}
\newcommand{\btriangle}{\raisebox{0.1pt}{\tikz{
			\draw[-,black,line width = 0.9pt](-2mm,0)--(2mm,0);
			\node[draw,scale=0.45,regular polygon, regular polygon sides=3,rotate=0,line width=1pt, fill=white]at (0,-0.25mm) {};}}\phantom{,}}
\newcommand{\bsquare}{\raisebox{-0.2pt}{\tikz{
			\draw[-,black,line width = 0.9pt](-2mm,0)--(2mm,0);
			\node[draw,scale=0.65,regular polygon, regular polygon sides=4,rotate=0,line width=1pt, fill=white]at (0,0) {};}}\phantom{,}}
\newcommand{\bdashed}{\raisebox{1.5pt}{\tikz{
			\draw[dash pattern={on 5.75pt off 2.5pt on 2.5pt off 2.5pt},black,line width = 0.9pt](-3.4mm,0)--(3.4mm,0);}}\phantom{,}}		
	
\title{Resolvent-based predictions for turbulent flow over anisotropic permeable substrates}
	

\author{A. Chavarin\aff{1},
    G. G\'omez-de-Segura\aff{2},
    R. Garc\'ia-Mayoral\aff{2},
    \and
    M. Luhar\corresp{\email{luhar@usc.edu}}\aff{1}
}
\affiliation{\aff{1}Department of Aerospace and Mechanical Engineering, University of Southern California, Los Angeles, California 90089 USA
\aff{2} Department of Engineering, University of Cambridge CB2 1PZ, UK
}

\maketitle

\begin{abstract}
Recent simulations indicate that streamwise-preferential porous materials have the potential to reduce drag in wall-bounded turbulent flows \citep{gomez-de-segura_garcia-mayoral_2019}.  This paper extends the resolvent formulation to study the effect of such anisotropic permeable substrates on turbulent channel flow. Under the resolvent formulation, the Fourier-transformed Navier-Stokes equations are interpreted as a linear forcing-response system. The nonlinear terms are considered the endogenous forcing in the system that gives rise to a velocity and pressure response. A gain-based decomposition of the forcing-response transfer function---the resolvent operator---identifies response modes (resolvent modes) that are known to reproduce important structural and statistical features of wall-bounded turbulent flows. The effect of permeable substrates is introduced in this framework using the Volume-Averaged Navier-Stokes equations and a generalized form of Darcy's law.  Substrates with high streamwise permeability and low spanwise permeability are found to suppress the forcing-response gain for the resolvent mode that serves as a surrogate for the energetic near-wall cycle. This reduction in mode gain is shown to be consistent with the drag reduction trends predicted by theory and observed in numerical simulations. Simulation results indicate that drag reduction is limited by the emergence of spanwise rollers resembling Kelvin-Helmholtz vortices beyond a threshold value of wall-normal permeability. The resolvent framework also predicts the conditions in which such energetic spanwise-coherent rollers emerge. These findings suggest that a limited set of resolvent modes can serve as the building blocks for computationally-efficient models that enable the design and optimization of permeable substrates for passive turbulence control. 
\end{abstract}

\section{Introduction}
Many flows of engineering and scientific interest involve permeable substrates.  The presence of such complex substrates can significantly alter the behavior of the near-wall turbulence.  A growing body of work suggests that appropriately-designed \textit{anisotropic} permeable substrates have the potential to suppress the dynamically-important near-wall (NW) cycle \citep{robinson1991coherent,waleffe1997self,jimenez_pinelli_1999} and reduce drag in wall-bounded turbulent flows \citep{hahn_je_choi_2002,itoh2006turbulent,abderrahaman-elena_garcia-mayoral_2017,gg_sharma_gm_2018,rosti2018turbulent,gomez-de-segura_garcia-mayoral_2019}.  Permeable materials can also be used to enhance turbulent mixing for applications in the development of high-efficiency thermal management systems and chemical reactors \citep{gad2007flow}.  Previous laboratory experiments and numerical simulations have provided significant insight into the effect of both isotropic and anisotropic permeable substrates on turbulent boundary layer and channel flows \citep[e.g.,][]{hahn_je_choi_2002, breugem2006influence, manes2011turbulent, zampogna_bottaro_2016, efstathiou2018mean, rosti2018turbulent, gomez-de-segura_garcia-mayoral_2019, kim2020experimental}.  For instance, it is well known that flows over porous materials are susceptible to a Kelvin-Helmholtz (KH) instability that gives rise to spanwise-coherent energetic rollers \citep{jimenez_pinelli_1999,breugem2006influence,efstathiou2018mean}.  The mechanism that could lead to drag reduction in turbulent flows over anisotropic materials is also reasonably well understood \citep{gomez-de-segura_garcia-mayoral_2019}.  Despite these advances, there are few reduced-complexity models that can be used to predict how a given porous substrate will affect the turbulent flow, i.e., whether it will suppress the NW cycle or give rise to KH rollers.  Given the vast parameter space available in the development of porous materials for passive flow control, such models can be useful tools for design and optimization.  In this study, we extend the resolvent analysis framework \citep{mckeon_sharma_2010,mckeon2017engine} to develop reduced-complexity models for turbulent flows over porous substrates.  We focus on evaluating the effect of anisotropic permeable materials that can give rise to drag reduction.  However, these models can also be used to evaluate the effect of porous materials for other applications or to provide insight into environmental flows over granular beds and vegetation canopies.

\subsection{Previous Work}
Recent numerical simulations show that streamwise-preferential permeable materials have the potential to yield as much as 25\% drag reduction in turbulent flows \citep{gomez-de-segura_garcia-mayoral_2019}. The physical mechanism underlying this drag reduction is similar to the mechanism that yields drag reduction over riblets \citep{walsh1984optimization,luchini_manzo_pozzi_1991,luchini1996,bechert1997experiments,garcia2011drag}.  Specifically, for anisotropic materials that have larger streamwise permeability than spanwise permeability, the streamwise mean flow penetrates into the substrate to a larger extent compared to the spanwise cross-flows arising from turbulent fluctuations \citep{luchini_manzo_pozzi_1991}. In other words, there is an offset between the virtual origins perceived by the mean flow and the turbulent fluctuations.  The virtual origin for the turbulent cross-flow can also be interpreted as the location for which the quasi-streamwise vortices associated with the NW cycle perceive a no-slip wall \citep{luchini1996,gomez-de-segura_garcia-mayoral_2019,gm_gs_fairhall_2019}.  Importantly, the offset in virtual origins for the mean streamwise flow and the turbulent cross-flows weakens the quasi-streamwise vortices associated with the NW cycle.  This reduces turbulent momentum transfer towards the substrate, which leads to a decrease in skin friction.  

Building on this concept, \citet{abderrahaman-elena_garcia-mayoral_2017} used the Brinkman equations to establish a relationship between the streamwise and spanwise permeabilities ($K^+_x,K^+_z$), and the streamwise and spanwise slip lengths ($\ell^+_x,\ell^+_z$), i.e., the lengths from the interface where the virtual wall would be perceived.  As shown in 
figure~\ref{fig:channel_dim}, $x$, $y$, and $z$ represent the streamwise, wall-normal, and spanwise coordinates, respectively.  A superscript $+$ denotes normalization with respect to the friction velocity, $u_\tau$, and viscosity, $\nu$.  The analysis of \citet{abderrahaman-elena_garcia-mayoral_2017} showed that $\ell^+_x\propto\kpxx$ and $\ell^+_z\propto\kpzz$ if the height of the substrate, $H$, is much larger than the permeability length scales $\kpxx$ and $\kpzz$.  The achievable drag reduction was estimated to be proportional to the difference between the streamwise and spanwise slip lengths, $\Delta D \propto \ell^+_x-\ell^+_z$, or equivalently, $\Delta D \propto \kpxx-\kpzz$.  This is consistent with the findings of \citet{busse_Sandham_2012}, who studied the effect of anisotropic slip length boundary conditions in turbulent channel flow simulations.

Recent direct numerical simulation (DNS) results obtained by \citet{gomez-de-segura_garcia-mayoral_2019} provide further support for the scaling developed by \citet{abderrahaman-elena_garcia-mayoral_2017}.  For these simulations, the Brinkman equations were used to model flow inside the permeable substrate and the Navier-Stokes equations were used in the fluid domain.  Stresses and velocities were matched at the interface between the permeable substrate and the unobstructed flow.  The permeable substrate was characterized by a permeability tensor of the form ${\bfK} =\diagK$, and the wall-normal permeability was set to be equal to the spanwise permeability, $K_y = K_z$.  The effect of substrate anisotropy was evaluated by systematically varying the streamwise and spanwise permeability, such that the anisotropy ratio $\phi_{xy}={\kpxx}/{\kpyy}$ ranged between $3.6$ and $11.4$. As predicted by \citet{abderrahaman-elena_garcia-mayoral_2017}, these simulations show that for surfaces for which permeability length scale is smaller than the size of the near wall turbulent structures, the initial decrease in drag is proportional to the difference in the slip length-scales, $\Delta D \propto \kpxx-\kpzz$. However, the simulations also show that the achievable drag reduction is limited by the appearance of energetic spanwise rollers resembling KH vortices.  Such rollers have been observed over isotropic permeable substrates \citep{breugem2006influence,efstathiou2018mean}, and they also contribute to performance degradation for riblets \citep{garcia2011hydrodynamic,garcia2012scaling}.  Linear stability analyses and simulation results show that the appearance of these spanwise rollers is controlled primarily by the wall-normal permeability \citep{abderrahaman-elena_garcia-mayoral_2017,gg_sharma_gm_2018,gomez-de-segura_garcia-mayoral_2019}.  Specifically, simulation results show that spanwise rollers become increasingly energetic as the wall-normal permeability exceeds $\kpyy \approx 0.4$.  The additional Reynolds shear stress produced by these rollers causes performance to deteriorate and ultimately leads to an increase in drag. 

These prior efforts show that the drag-reducing performance of anisotropic permeable substrates is dictated by two key factors: the suppression of near-wall cycle and the emergence of energetic spanwise-coherent rollers.  The physically-motivated slip length arguments presented earlier provide useful insight into the first effect.  These arguments predict that the initial decrease in drag is proportional to the difference between the streamwise and spanwise permeability length scales, $\Delta D \propto \kpxx-\kpzz$.  However, it is unclear if this relationship also holds for more complex surfaces that are not characterized by diagonal permeability tensors of the form ${\bfK} =\diagK$.  Moreover, these slip length arguments are based on solutions to the Brinkman equations in the porous medium, which are coupled to the fluid domain via velocity and stress-matching boundary conditions at the fluid-substrate interface.  Recent studies show that the interfacial boundary conditions may be better characterized by a slip-length tensor \citep{lacis_Bagheri_2017,lacis_Sudhakar_Pasche_Bagheri_2020,bottaro_2019}.  The effect of such slip-length models on the near-wall turbulence remains to be studied.  Similarly, linear stability analyses are able to predict the emergence of spanwise-coherent KH rollers over permeable substrates as the wall-normal permeability increases.  However, such analyses fail to accurately predict the exact threshold for $\kpyy$ beyond which KH rollers become energetic \citep{abderrahaman-elena_garcia-mayoral_2017,gg_sharma_gm_2018}.  Further, the streamwise wavelengths predicted to be most unstable do not match the length scale of the spanwise rollers observed in simulations.  

\subsection{Contribution and Outline}
In this study, we develop a reduced-order modeling framework grounded in resolvent analysis \citep{mckeon_sharma_2010,mckeon2017engine} that can be used to predict the effect of substrates with known permeability on the NW cycle and test for the emergence of KH rollers.  Under the resolvent formulation, the Navier-Stokes equations are interpreted as a forcing-response system in which the nonlinear convective terms are treated as the forcing to the linear system that generates a velocity and pressure response.  A gain-based decomposition of the resolvent operator, which is the linear transfer function that maps the nonlinear forcing to the velocity and pressure response, is used to identify high-gain forcing and response modes across spectral space. Specific high-gain response modes (\textit{resolvent modes}) have been shown to serve as useful models for dynamically-important flow features such as the NW cycle \citep{moarref2013model}.  This means that, as a starting point, the effect of any control can be evaluated on these individual resolvent modes instead of the full turbulent flow field.  Indeed, previous studies show that the gain for the NW resolvent mode is a useful predictor of drag reduction performance for both active \citep{luhar2014opposition,nakashima2017assessment,toedtli2019predicting} and passive \citep{luhar2015framework,chavarin2020resolvent} control techniques in wall-bounded turbulent flows.  In particular, recent work by \citet{chavarin2020resolvent} shows that the gain for the NW resolvent mode is a useful surrogate for total drag reduction in turbulent flows over riblets.  Riblet geometries that lead to drag reduction in experiments and high-fidelity simulations are found to reduce the forcing-response gain for the NW resolvent mode relative to its smooth wall value.  In addition, \citet{chavarin2020resolvent} show that the resolvent framework is also able to predict the emergence of spanwise rollers resembling KH rollers over certain riblet geometries, which is consistent with previous DNS results \citep{garcia2011hydrodynamic}.  Motivated by these prior modeling successes, we consider the effect of anisotropic porous substrates on resolvent modes resembling the NW cycle and spanwise-coherent structures resembling KH rollers. To enable a direct comparison with the simulation results of \citet{gomez-de-segura_garcia-mayoral_2019}, we consider a symmetric channel geometry at friction Reynolds number $\Ret = 180$ and substrates with identical wall-normal and spanwise permeabilities, i.e., substrates with $\phi_{yz}=\kpyy/\kpzz=1$.
We model the flow in the substrate using the volume-averaged Navier-Stokes equations, in which the effect of the permeable substrate is included via a permeability tensor.  However, this modeling framework can be extended to include more sophisticated interfacial boundary conditions \citep{lacis_Bagheri_2017,lacis_Sudhakar_Pasche_Bagheri_2020}, and to account for inertial effects via the so-called Forchheimer term \citep{breugem2006influence,whitaker2013method}.

The remainder of this paper is structured as follows.  Section~\ref{sec:methods} describes the permeable substrate model used here, the extended resolvent formulation, as well as the numerical methods used for resolvent analysis.  Section~\ref{sec:results} presents model predictions for the effect of anisotropic permeable substrates on the NW resolvent mode as well as spanwise-coherent resolvent modes resembling KH rollers.  These predictions are compared against DNS results from \citet{gomez-de-segura_garcia-mayoral_2019}.  We also pursue a limited sensitivity analysis of model predictions to the exact form of the mean profile used to construct the resolvent operator. Specifically, we compare predictions made using a synthetic mean profile that is computed using an eddy viscosity model against predictions generated using the mean velocity profiles obtained in DNS by \citet{gomez-de-segura_garcia-mayoral_2019}.  Section~\ref{sec:conclusion} concludes the paper. 

\section{Methods}\label{sec:methods}
In this section, we present the equations used to model flow in the porous medium (\S\ref{sec:gov_mods}), briefly describe the resolvent formulation and discuss its extension to account for permeable substrates (\S\ref{sec:resolvent_analysis}), present the boundary conditiobluns imposed at the fluid-substrate interface (\S\ref{sec:boundary_conds}), discuss the mean velocity profiles used to construct the resolvent operator (\S\ref{sec:mean_vel_methods}), and describe the numerical method used to implement the analysis (\S\ref{sec:numerical_methods}). 

\subsection{Accounting for Permeable Substrates}\label{sec:gov_mods}

The resolvent framework is reformulated using the volume-averaged Navier-Stokes equations.  Volume-averaging gives rise to two additional terms: a term representing the sub-filter scale stresses and a surface filter term that accounts for the force exerted by the solid phase of the permeable medium on the fluid phase \citep{Whitaker_1969,Whitaker_1996,whitaker2013method}.  A typical closure model for the surface filter term involves parameterizing the flow resistance using the Darcy permeability tensor and the so-called Forchheimer correction term that accounts for inertial effects \citep{Whitaker_1996}.  
This model has been used in previous numerical simulations of flow over porous substrates \citep{breugem2006influence,rosti2015direct,rosti2018turbulent} as well as in linear stability analyses \citep{tilton_cortelezzi_2006,tilton_cortelezzi_2008}.  The volume-averaged Navier-Stokes equations and continuity constraint for flow through a porous medium with porosity $\epsilon$, dimensionless permeability $\bfK$, and dimensionless Forchheimer resistance $\mathsfbi{F}$ can be expressed as:
\begin{subequations}\label{eqn:governing_eq}
\begin{equation} 
	\pdv{\bbfu}{t} + \frac{1}{\epsilon}\grad\cdot\big(\epsilon\bbfu\bbfu+\epsilon\boldsymbol{\tau} \big) = -\frac{1}{\epsilon}\grad(\epsilon \langle p\rangle) + \frac{1}{\epsilon\Ret}\grad^2 (\epsilon\bbfu)-\frac{\epsilon}{\Ret}\bfK^{-1}(\mathsfbi{I}+\mathsfbi{F})\bbfu 
\end{equation}
and
\begin{equation}
	\grad\cdot(\epsilon\bbfu) = 0.
\end{equation}
\end{subequations}
In the equations above, $\langle\cdot\rangle$ represents a volume-averaged quantity, $\bbfu$ is the dimensionless velocity, $\langle p \rangle$ is the dimensionless pressure, and $t$ is dimensionless time.  The equations presented above have been normalized using the channel half-height $h$ and the friction velocity $u_\tau$.  The friction Reynolds number is given by $\Ret=u_{\tau}h/\nu$ and the dimensionless permeability defined as $\bfK = \bfK^\dagger/h^2$, where $\bfK^\dagger$ is the dimensional permeability.  The quantity $\boldsymbol{\tau}=\langle\bfu\bfu\rangle-\bbfu\bbfu$ represents the sub-filter scale stresses which arise from volume averaging the Navier-Stokes equations.

For our analysis we consider the following simplifications.  Consistent with prior numerical simulations \citep{rosti2018turbulent,gomez-de-segura_garcia-mayoral_2019}, we focus on substrates characterized by a permeability tensor of the form ${\bfK} =\diagK$ with the ratio of the wall-normal and spanwise permeabilities set to unity, i.e., $K_y=K_z$. Note that the permeability tensor is symmetric, and so an eigenvalue decomposition can be used to identify its principal values and directions (or axes).  The assumed form of the permeability tensor implies that its principal directions align with the streamwise, wall-normal, and spanwise directions of the flow. Second, we omit the nonlinear Forchheimer correction term, $\bf{F}$, that is used to account for inertial effects in flows through porous media \citep{ochoa1995momentum1,ochoa1995momentum2,breugem2006influence,whitaker2013method}. This assumption is again consistent with the numerical simulations we will use to test model predictions. Third, we assume that the porosity of permeable substrates is $\epsilon\approx1$ to maximize any potential drag reduction \citep{abderrahaman-elena_garcia-mayoral_2017}.  Finally, since we are primarily interested in structures that are much larger than the characteristic length scale of the porous medium (i.e., NW cycle and KH rollers), we neglect the sub-filter scale stresses \citep{breugem2006influence}.  With these assumptions, (\ref{eqn:governing_eq}a) and (\ref{eqn:governing_eq}b) can be expressed as:
\begin{subequations}\label{eqn:governing_eq1} 
\begin{equation}
    \pdv{\bfu}{t} + \grad\cdot\big(\bfu\bfu \big) = -\grad(p) + \frac{1}{\Ret}\grad^2 \bfu -\frac{1}{\Ret}\bfK^{-1}\bfu  
\end{equation}
and
\begin{equation}
    \grad\cdot\bfu = 0,   
\end{equation}
\end{subequations}
where the $\langle\cdot\rangle$ notation has been eliminated for simplicity. The unobstructed fluid domain is characterized by infinite permeability.  In this region, the permeability term goes to zero and (\ref{eqn:governing_eq1}a) reduces to the standard Navier-Stokes momentum equation.  

\begin{figure}
	\centering
	\includegraphics[scale=0.8]{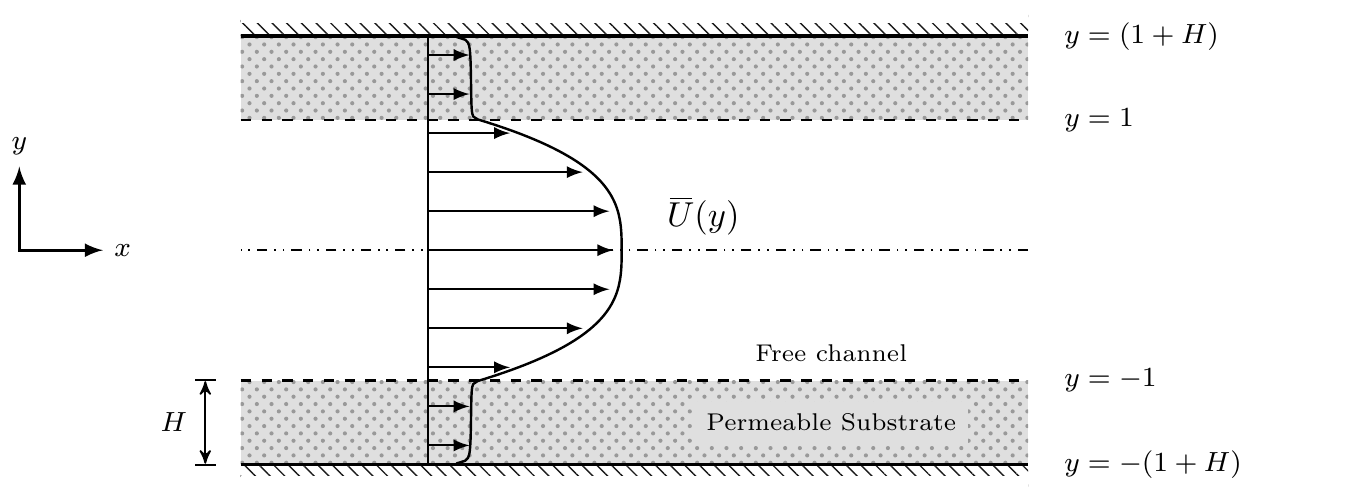}
	\caption{Schematic showing the symmetric channel flow configuration considered in this paper. 		
	\label{fig:channel_dim}}. 
\end{figure}

Figure~\ref{fig:channel_dim} shows the symmetric channel flow configuration considered in this study.  The unobstructed region corresponds to $y\in [-1,1]$.  The regions occupied by the permeable substrates correspond to $y\in [-(1+H),\,-1)$ and $y\in (1,1+H]$.  The height of the permeable substrate is $H$.  Note that all lengths are normalized by the channel half-height. 
	
\subsection{Modified Resolvent Analysis}\label{sec:resolvent_analysis}
The resolvent formulation for wall-bounded turbulent flows proposed by \citet{mckeon_sharma_2010}---and employed in several recent flow control studies \citep{luhar2014opposition,luhar2015framework,nakashima2017assessment,toedtli2019predicting,chavarin2020resolvent}---is extended to account for the presence of permeable substrates as follows. For an extended discussion of resolvent analysis and its applications, the reader is referred to \citet{mckeon2017engine}. 

Construction of the modified resolvent operators begins with a standard Reynolds-decomposition of the simplified volume-averaged Navier-Stokes equations in (\ref{eqn:governing_eq1}).  The velocity and pressure fields are decomposed into a time-averaged component (denoted by $\overbar{(\cdot)}$) and a fluctuating component about this average (denoted by ${(\cdot)^\prime}$).  Under this decomposition, the velocity field is expressed as $\bfu(t,{\bf{x}})=\bfU({\bf{x}})+\bfu^\prime(t,{\bf{x}})$ and the pressure field is expressed as $p(t,{\bf{x}}) = \overbar{P}({\bf{x}}) + p^\prime(t,{\bf{x}})$.  Note that $\bfU({\bf{x}}) = [\mnU(y),0,0]^T$ represents the turbulent mean profile.  Next, the velocity and pressure fluctuation are Fourier-transformed in the homogeneous streamwise and spanwise directions as well as in time:
\begin{equation}
\begin{bmatrix}
\bfu^\prime(t,\bf{x}) \\
p^\prime(t,\bf{x})
\end{bmatrix} 
= \scalerel{\iiint}{\begin{bmatrix}
\bfu_{\bkv}(y) \\
p_{\bkv}(y)
\end{bmatrix}}\exp(-\mathrm{i}\omega t+\mathrm{i}\kxx x +\mathrm{i}\kzz z)\dd{\omega}\dd{\kxx}\dd{\kzz}.
\end{equation}
In the expression above, $\kxx$ is the streamwise wavenumber, $\kzz$ is the spanwise wavenumber, and $\omega$ is the frequency.  The Fourier coefficients for the velocity and pressure field at a given wavenumber-frequency combination, $\bkv = (\kxx,\kzz,\omega)$, are denoted $\bfu_{\bkv}$ and $p_{\bkv}$.  Under the Fourier transform, the equations in (\ref{eqn:governing_eq1}) can be expressed compactly as:
\begin{equation}
\begin{bmatrix} 
{\bfu}_{\bkv} \\  {p}_{\bkv}
\end{bmatrix} 
= \Bigg(-\mathrm{i}\omega \mqty[\dmat{\mathsfbi{I},0}] -\mqty[\boldsymbol{\mathcal{L}}_{\bkv} & -\tnabla \\ -\tnabla^T & 0]\Bigg)^{-1} \mqty[\mathsfbi{I} \\ 0]\boldsymbol{f}_{\bkv} = \boldsymbol{\mathcal{H}}_{\bkv}\boldsymbol{f}_{\bkv}. \label{eqn:mod_resolvent}
\end{equation}
In the expression above, the first row represents the momentum equations and the second row represents the continuity constraint.  The operator 
\begin{equation}
\setlength{\arraycolsep}{-3pt}
\renewcommand{\arraystretch}{1.5}
\boldsymbol{\mathcal{L}}_{\bkv} = \begin{bmatrix}
-\mathrm{i}\kxx \mnU + \Ret\inv(\tnabla^2 -\bfK\inv)  & -\dv{\mnU}{y}  & 0 \\
0     &  -\mathrm{i}\kxx \mnU + \Ret\inv(\tnabla^2 -\bfK\inv ) & 0 \\
0     &          0                        &  -\mathrm{i}\kxx \mnU + \Ret\inv(\tnabla^2 - \bfK\inv) 
\end{bmatrix}   \label{eqn:that_l_matrix}\\
\end{equation}

represents the linear dynamics in (\ref{eqn:governing_eq1}), and $\boldsymbol{f}_{\bkv}$ is the Fourier coefficient for the non-linear terms.  The differential operator $\tnabla$ is defined as $\tnabla=(\mathrm{i}\kxx,\pdv*{}{y},\mathrm{i}\kzz)$, and so the symbols $\bf\tnabla^T$  and  $\bf\tnabla$ essentially represent the Fourier-transformed divergence and gradient operators, respectively. Similarly, the Laplacian is defined as $\tnabla^2=(-\kxx^2-\kzz^2+\pdv*[2]{}{y})$.  Note that the velocity and pressure response at a given wavenumber-frequency combination constitute a traveling wave flow field with streamwise wavelength $\lambda_x=2\pi/\kxx$ and spanwise wavelength $\lambda_z=2\pi/\kzz$ that is moving downstream at speed $c=\omega/\kxx$.  The transfer function that maps the nonlinear forcing $\boldsymbol{f}_{\bkv}$ to the velocity and pressure response $[\bfu_{\bkv},{p}_{\bkv}]^T$ in (\ref{eqn:mod_resolvent}) is the resolvent operator, $\mathcal{H}_{\bkv}$.  The central difference between this resolvent operator for channel flow over permeable substrates and its smooth wall counterpart is the inclusion of the Darcy permeability tensor $\bfK$ in (\ref{eqn:that_l_matrix}). Note that the resistance exerted by the Darcy term is linear with respect to $\bfu$.

As detailed in prior studies \citep{mckeon_sharma_2010,moarref2013model,luhar2014opposition,luhar2015framework}, a singular value decomposition (SVD) of the discretized resolvent operator (see Section~\ref{sec:numerical_methods}) is used to identify a set of orthonormal forcing and response modes that are ordered based on their forcing-response gain under an $L^2$ energy norm.  To enforce this norm, the discretized resolvent operator in (\ref{eqn:mod_resolvent}) is scaled as follows:  
\begin{subequations}\label{eqn:weighted_resolvent}
	\begin{equation}
	\begin{bmatrix} \WmatU& 0 \end{bmatrix}
	\begin{bmatrix}
	\bfu_{\bkv} \\  {p}_{\bkv}
	\end{bmatrix} = 
	\left(\begin{bmatrix} \WmatU& 0 \end{bmatrix}
	\boldsymbol{\mathcal{H}}_{\bkv} \WmatF\inv\right) \WmatF\boldsymbol{f}_{\bkv}
	\end{equation}
	or
	\begin{equation}
	\WmatU\bfu_{\bkv} = \boldsymbol{\mathcal{H}}^w_{\bkv} \WmatF\boldsymbol{f}_{\bkv}.
	\end{equation}
\end{subequations}
Here, $\WmatU$ and $\WmatU$ incorporate numerical quadrature weights for the entire domain spanning $y\in [-(1+H),(1+H)]$.  With this weighting, the SVD of the scaled resolvent:
\begin{subequations}\label{eqn:umod_resolvent}
\begin{equation}
\boldsymbol{\mathcal{H}}^w_{\bkv}= \sum_{m} \psi_{\bkv,m} \sigma_{\bkv,m} \phi^{\ast}_{\bkv,m},
\end{equation}
where
\begin{equation}
     \sigma_{\bkv,1}>\sigma_{\bkv,2}>...>0, \quad \psi_{\bkv,l}^*\psi_{\bkv,m} = \delta_{lm}, \quad \phi_{\bkv,l}^*\phi_{\bkv,m} = \delta_{lm}
\end{equation}
\end{subequations}
yields forcing modes $\boldsymbol{f}_{\bkv,m}=\WmatF\inv\phi_{\bkv,m}$ and velocity responses $\bfu_{\bkv,m}={\WmatU\inv}\psi_{\bkv,m}$ that have unit energy when integrated across the entire domain spanning $y\in [-(1+H),(1+H)]$.  In other words, this scaling ensures that
\begin{equation}\label{eqn:SVDNorm2}
	\int_{-(1+H)}^{(1+H)} \bfu^*_{\bkv,l} \bfu^{}_{\bkv,m} \,dy = \delta_{lm},\quad \int_{-(1+H)}^{(1+H)} \boldsymbol{f}^*_{\bkv,l} \boldsymbol{f}^{}_{\bkv,m} \,dy = \delta_{lm}.
\end{equation}
In (\ref{eqn:umod_resolvent})-(\ref{eqn:SVDNorm2}), the superscript $(\cdot)^*$ denotes a conjugate transpose. 

A major contribution of the resolvent framework lies in the finding that the forcing-response transfer function tends to low-rank at $\bkv$ combinations that are energetic in wall-bounded turbulent flows.  Often, the first singular value is an order of magnitude larger than subsequent singular values, $\sigma_{\bkv,1} \gg \sigma_{\bkv,2}>...$, and so the resolvent operator can be well approximated using a rank-1 truncation after the SVD \citep{mckeon_sharma_2010,moarref2013model}:
\begin{equation}\label{eqn:rank1}
    \boldsymbol{\mathcal{H}}^w_{\bkv} \approx \psi_{\bkv,1} \sigma_{\bkv,1} \phi^{\ast}_{\bkv,1}.
\end{equation}
The expressions in (\ref{eqn:weighted_resolvent})-(\ref{eqn:rank1}) show that forcing in the direction of the first forcing mode $\boldsymbol{f}_{\bkv,1}=\WmatF\inv\phi_{\bkv,1}$ generates a velocity response $\bfu_{\bkv,1}={\WmatU\inv}\psi_{\bkv,1}$ that is amplified by factor $\sigma_{\bkv,1}$.  Put another way, a forcing of the form $\boldsymbol{f}_{\bkv,1}$ to the unscaled resolvent operator in (\ref{eqn:mod_resolvent}) generates a velocity and pressure response $\sigma_{\bkv,1}[\bfu_{\bkv,1},p_{\bkv,1}]^T$.  Under the $L^2$ scaling used here, $\sigma_{\bkv,1}^2$ is a measure of energy amplification. 

Recent modeling efforts for active and passive flow control techniques show that specific rank-1 modes serve as useful surrogates for the dynamically-important NW cycle.  Specifically, the ability of a control technique to suppress the forcing-response gain for modes with wavenumber-frequency combinations corresponding to $\llx \approx 10^3$, $\llz \approx 10^2$, and $c^+ \approx 10$ (i.e., similar to the length and velocity scales associated with NW streaks) has been shown to be a useful predictor of drag reduction performance \citep{luhar2014opposition,nakashima2017assessment,chavarin2020resolvent}.  Building on these prior efforts, in this study we evaluate the effect of anisotropic permeable substrates on the rank-1 resolvent mode that serves as a surrogate for the NW cycle. A reduction in gain for this mode relative to the smooth-wall value is interpreted as mode suppression, which is indicative of drag reduction.  In addition, we also test whether the permeable substrates lead to an increase in principal singular values for spanwise-coherent modes (e.g., with $\kzz = 0$) that resemble KH rollers. Since we only consider the rank-1 truncation shown in (\ref{eqn:rank1}) for the remainder of this paper, we drop the additional subscript $1$ to simplify notation.

\subsection{Boundary and Interface Conditions}\label{sec:boundary_conds}
As discussed in Section~\ref{sec:numerical_methods} below, the resolvent operator is discretized using spectral discretization and rectangular block matrices as described by \citet{block_operators_aurentz_trefethen}. This approach enables us to use two different sets of equations in the unobstructed region and porous domain (i.e., without and with the permeability term), and couple the two via appropriate interfacial conditions. 
As shown in figure~\ref{fig:channel_dim}, the unobstructed channel corresponds to the region corresponding to $y\in [-1,1]$ and the upper and lower permeable regions correspond to $y\in (1,1+H]$ and $y\in [-(1+H),-1)$, respectively.  At the lower and upper substrate walls, $y=\pm (1+H)$, we apply no-slip boundary conditions.  At the interfaces between the porous medium and the unobstructed flow, $y = \pm 1$, we impose continuity in all three components of velocity and pressure. We also impose continuity in the streamwise and spanwise shear at the interface.  These boundary conditions can be summarized as follows:
\begin{subequations}\label{eqn:bndry_cond}
	\begin{equation}
	\bfu = 0 \quad\text{at}\quad y = \pm (1+H),
	\end{equation}
	\begin{equation}
	\left.\bm{u}\right\rvert_{y_+} =\left.\bm{u}\right\rvert_{y_-}\quad\text{and}\quad \left. p \right\rvert_{y_+} =\left. p \right\rvert_{y_-} \quad\text{at}\quad y = \pm 1,
	\end{equation}
	\begin{equation}
	\left.\pdv{{u}}{y}\right\rvert_{y_+} = \ieps\left.\pdv{{u}}{y}\right\rvert_{y_-} \quad\text{and}\quad \left.\pdv{{w}}{y}\right\rvert_{y_+} = \ieps\left.\pdv{{w}}{y}\right\rvert_{y_-} \quad \text{at} \quad y = -1,
	\end{equation}
	\begin{equation}
	\left.\pdv{{u}}{y}\right\rvert_{y_-} = \ieps\left.\pdv{{u}}{y}\right\rvert_{y_+} \quad\text{and}\quad \left.\pdv{{w}}{y}\right\rvert_{y_-} =\ieps\left.\pdv{{w}}{y}\right\rvert_{y_+} \quad \text{at} \quad y = 1.\hphantom{-}
    \end{equation}
\end{subequations}
In the expressions above, $y_+$ and $y_-$ denote values taken on either side of the porous substrate-unobstructed flow interface.  Note that the boundary conditions shown in (\ref{eqn:bndry_cond}c,d) do not penalize momentum transfer into the permeable substrate.  These conditions also assume that the effective viscosity in the porous medium is $\tilde{\nu} = \nu/\epsilon$.  Per the discussion presented in \citet{abderrahaman-elena_garcia-mayoral_2017} and \citet{gomez-de-segura_garcia-mayoral_2019}, these assumptions are reasonable for a highly-connected medium with high porosity, $\epsilon \approx 1$.  For a poorly-connected medium, previous studies show that a stress \textit{jump} boundary condition may be more appropriate, and that more complex effective viscosity models may be needed \citep[see e.g.,][]{ochoa1995momentum1,Minale_2014}.
Also keep in mind that the equations above imply a sharp transition between the porous medium and the unobstructed fluid. Previous studies employing the volume-averaged Navier-Stokes equations have typically assumed the existence of a finite transition zone between the homogeneous porous and homogeneous fluid regions \citep{ochoa1995momentum1,breugem2006influence}.

Finally, since we are interested in permeable substrates that have the highest potential for drag reduction, we assume a porosity of $\epsilon=1$.  For this value of porosity, the boundary conditions shown in (\ref{eqn:bndry_cond}) are similar to those used in previous high-fidelity simulations \citep{gomez-de-segura_garcia-mayoral_2019}.  Emulation of the channel configuration and boundary conditions used by \citet{gomez-de-segura_garcia-mayoral_2019} allows for a direct comparison between model predictions and simulation results. 
	
\subsection[mean velocity profiles]{Mean Velocity Profile}\label{sec:mean_vel_methods}
As shown in (\ref{eqn:that_l_matrix}), construction of the resolvent operator requires knowledge of the mean velocity profile $\mnU(y)$.  Here, we use two different sets of mean velocity profiles to generate model predictions: (i) mean profiles obtained in DNS by \citet{gomez-de-segura_garcia-mayoral_2019}, and (ii) mean profiles generated using a synthetic eddy viscosity profile.  The DNS mean profile dataset consists of 22 different cases: 8 profiles for $\phi_{xy} = 3.6$; 7 profiles for $\phi_{xy} = 5.5$; and 7 profiles for $\phi_{xy} = 11.4$. For all 22 cases tested in DNS, the ratio between the wall-normal and spanwise permeability length scales was $\phi_{yz}= \kpyy/\kpzz =1$. Each of the DNS profiles contained 153 points spanning the unobstructed region, $y\in [-1,1]$.  Within the permeable medium, the analytical solutions developed by \citet{gomez-de-segura_garcia-mayoral_2019} were used.  These mean profiles were interpolated onto our Chebyshev grid using the modified Akima interpolation algorithm.  

For each of the 22 different DNS cases, the synthetic profiles were generated as follows. The modified governing equation (\ref{eqn:governing_eq1}) was Reynolds-averaged to yield the following equation for the streamwise mean flow:  
\begin{equation}\label{eqn:mean_eq}
	\frac{1}{\Ret}\dv[2]{\mnU}{y}-\frac{\mnU}{\Ret K_x} -\frac{d(\overbar{u^\prime v^\prime})}{dy} =\frac{d\overbar{p}}{dx},
\end{equation} 
where $K_x$ is the streamwise component of the permeability tensor. The Reynolds shear stress term in (\ref{eqn:mean_eq}) was estimated using a modified version of the eddy viscosity formulation proposed by \citet{reynolds_tiederman_1967}:
\begin{equation}
	\nu_e =\left\lbrace
	\begin{array}{ll}
	\frac{1}{2}\left\lbrack 1+\left\{\dfrac{c_2\Ret}    {3}(2y-y^2)(3-4y+2y^2)\right.\right. &\\
	\qquad\qquad\times  \left.\left.\left(1-\exp((|y-1|-1)\dfrac{\Ret}{c_1})\right)\right\}^2
	\right\rbrack^{1/2}- \frac{1}{2} & \abs{y} \le 1, \\
	\quad 0 & \abs{y} > 1.
	\end{array}\right.\label{eqn:eddy_viscosity}
\end{equation}

Here, the eddy viscosity has been normalized by the fluid viscosity $\nu$. To generate the synthetic profiles, we used the values $c_1 = 46.2$ and $c_2 = 0.61$, which were identified by \citet{moarref2012model} as yielding the best fit to the mean profiles obtained in smooth wall DNS at $Re_{\tau} \approx 180$. Thus, the Reynolds shear stress term was modeled using a standard smooth-wall eddy viscosity profile in the unobstructed region of the channel and assumed to be zero in the permeable substrate.  In other words, this model assumes that there is no turbulence penetration into the porous medium.  In reality, the penetration of turbulent cross-flows into the permeable substrate will depend on the spanwise and wall-normal components of permeability, and so the model above is only valid for small $K_y$ and $K_z$. Finally, equations (\ref{eqn:mean_eq}) and (\ref{eqn:eddy_viscosity}) were combined to yield the following equation for the mean velocity profile:
\begin{equation}\label{eqn:mean_eq2}
	\frac{1}{\Ret}\left((1+\nu_e)\dv[2]{}{y} + \dv{\nu_e}{y}\dv{}{y}-\frac{1}{K_x}\right)\mnU =\frac{d\overbar{p}}{dx},
\end{equation} 
which was solved numerically to yield $\mnU(y)$.  

We recognize that the procedure outlined above to estimate the synthetic mean profile involves significant assumptions.  However, as we show below, the interfacial slip velocities generated using this model are in good agreement with the DNS mean profiles in the initial drag reduction regime over anisotropic permeable substrates, i.e., until the point of performance degradation.  Moreover, the resolvent-based predictions obtained using these profiles are also similar to those computed using the DNS mean profiles.

\subsection{Numerical Methods}\label{sec:numerical_methods}
The resolvent operator and the equations used to synthesize the mean velocity profile are discretized in the wall-normal direction using spectral discretization methods involving rectangular block operators, as described by \citet{block_operators_aurentz_trefethen}. Each differential operator is discretized using Chebyshev polynomials and the resulting matrices are rectangular. The size of these matrices is $[N \times N+n]$ and where $n$ represents the dimension of the operator null space. The overall block operator is made square by appending $n$ boundary conditions. By using these block operators, we can deal with a system of boundary value problems coupled through boundary conditions. 
For a more thorough discussion of these methods readers are directed to \citet{block_operators_aurentz_trefethen}. 

As shown in figure~\ref{fig:channel_dim}, for our problem configuration the channel is separated into 3 regions: the lower permeable substrate spanning $y \in [-H,-1)$, the free channel spanning $y \in [-1,1]$, and the upper permeable substrate spanning $y \in (1,H]$.  Although the configuration is symmetric across the channel centerline, we consider the entire channel to retain resolvent modes that are both symmetric and anti-symmetric across the channel centerline \citep{moarref2013model,luhar2015framework}.  Each of the three regions is discretized using $N$ Chebyshev points and so the total number of Chebyshev points for the channel is $3N$. A grid convergence study showed that the normalized change in singular values is of $\mathcal{O}(10^{-6})$ for grid sizes $N=80$, $N=112$, and $N=200$. Each resolvent mode computation takes approximately 0.7s for $N=80$, approximately 2.0s for $N=112$ and 7.0s for $N=200$. The results presented below were generated using $N=112$, which corresponds to a total of $3N=336$ grid points across the the entire channel.
	
\section{Results and Discussion}\label{sec:results}
In this section, we compare resolvent-based predictions for the NW mode (Section~\ref{sec:nw_resolvent_mode}) and the emergence of energetic spanwise rollers (Section~\ref{sec:KH_rollers}) against DNS observations. Before that, we briefly compare the mean velocity profiles obtained from DNS against the synthetic profiles generated using the eddy viscosity model (Section~\ref{sec:mean_vel_results}).

\subsection{Mean Velocity Profiles}\label{sec:mean_vel_results}

\begin{figure}
	\centering
	\includegraphics[scale=0.8]{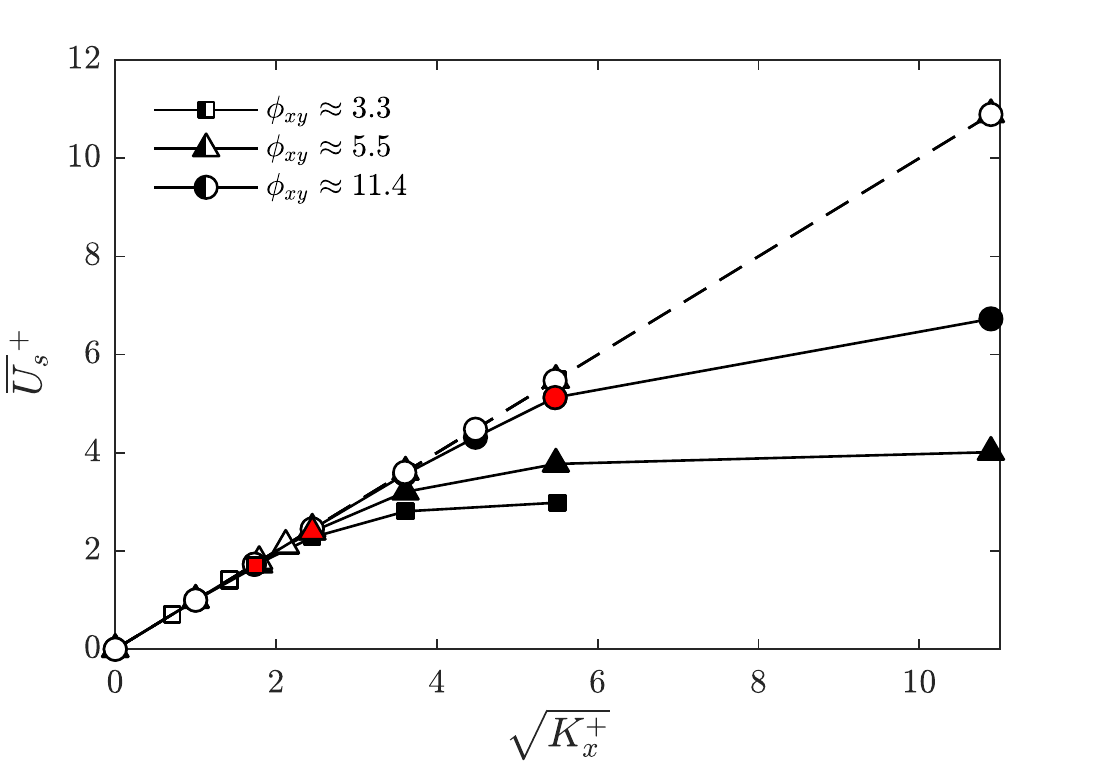}
	\caption{Predicted slip velocity at the porous interface as a function of streamwise permeability $\protect\kpxx$. Closed symbols show DNS results from \citet{gomez-de-segura_garcia-mayoral_2019}. Open symbols show predictions made using (\ref{eqn:mean_eq2}). Red symbols indicate the cases for which wall-normal permeability first exceeds $\protect\kpyy \gtrsim 0.4$.  This is roughly the threshold above which drag reduction performance deteriorates in DNS.}\label{fig:slip_figure} 
\end{figure}

Figure~\ref{fig:slip_figure} compares the interfacial slip velocity $\mnU_s^+$ predicted using the synthetic eddy viscosity profile (\ref{eqn:eddy_viscosity}) against those obtained in DNS by \citet{gomez-de-segura_garcia-mayoral_2019} for each of the 22 different cases being considered here, with anisotropy ratios $\phi_{xy} = 3.6$, $5.5$, and $11.4$.  The slip velocities for the synthetic profiles all collapse together onto a straight line corresponding to $\mnU_s^+ \approx \kpxx$.  In other words, the slip velocity only depends on the streamwise component of permeability, which is consistent with the assumptions outlined in Section~\ref{sec:mean_vel_methods}.  The synthetic mean profiles do not account for the effect of wall-normal or spanwise permeability, which are likely to determine the extent to which turbulence penetrates into the permeable substrate.

The synthetic slip velocities (open symbols) show close agreement with DNS results (closed symbols) for low streamwise permeabilities but begin to deviate from DNS results at higher $\kpxx$.  Moreover, the threshold values of $\kpxx$ above which the synthetic predictions begin to deviate from DNS results depend on the anisotropy ratio, $\phi_{xy}$. As an example, for $\phi_{xy} = 3.6$, synthetic $\mnU_s^+$ begin deviating from DNS results for $\kpxx \gtrsim 2$.  For $\phi_{xy} = 11.4$, the predicted slip velocities deviate significantly from DNS results for $\kpxx \gtrsim 6$.  Closer inspection of this trend shows that the synthetic slip velocities deviate from DNS data above a constant threshold value for the wall-normal permeability for all anisotropy ratios, $\kpyy \gtrsim 0.4$.  This value of $\kpyy$ corresponds closely to the conditions in which Kelvin-Helmholtz-type rollers appear over the permeable substrates in the simulations of \citet{gomez-de-segura_garcia-mayoral_2019}. The appearance of these rollers is likely to generate significant interfacial turbulence that penetrates the porous medium. The eddy viscosity model used to generate the synthetic mean profiles does not account for these effects. The eddy viscosity model shown in (\ref{eqn:eddy_viscosity}) assumes that no turbulence penetrates into the porous medium, and that the turbulence in the unobstructed region remains similar to that over a smooth wall regardless of the value of $\kpyy$ and $\kpzz$.

For cases in which the interfacial slip velocities agree, the synthetic mean profiles are similar to the DNS profiles across the entire channel (data not shown here for brevity). 
For these cases, resolvent-based predictions are not very sensitive to the choice of mean profile.  Moreover, as we show in Section~\ref{sec:KH_rollers}, resolvent analysis is able to predict the emergence of high-gain KH rollers over permeable substrates as the wall-normal permeability increases beyond $\kpyy \gtrsim 0.4$.  These observations indicate that resolvent analysis carried out using the synthetic mean profile can be a useful design tool for the design of passive flow control using anisotropic permeable substrates.

\subsection{Near-Wall Resolvent Mode}\label{sec:nw_resolvent_mode}

\begin{figure}
	\centering
	\includegraphics[width=\textwidth]{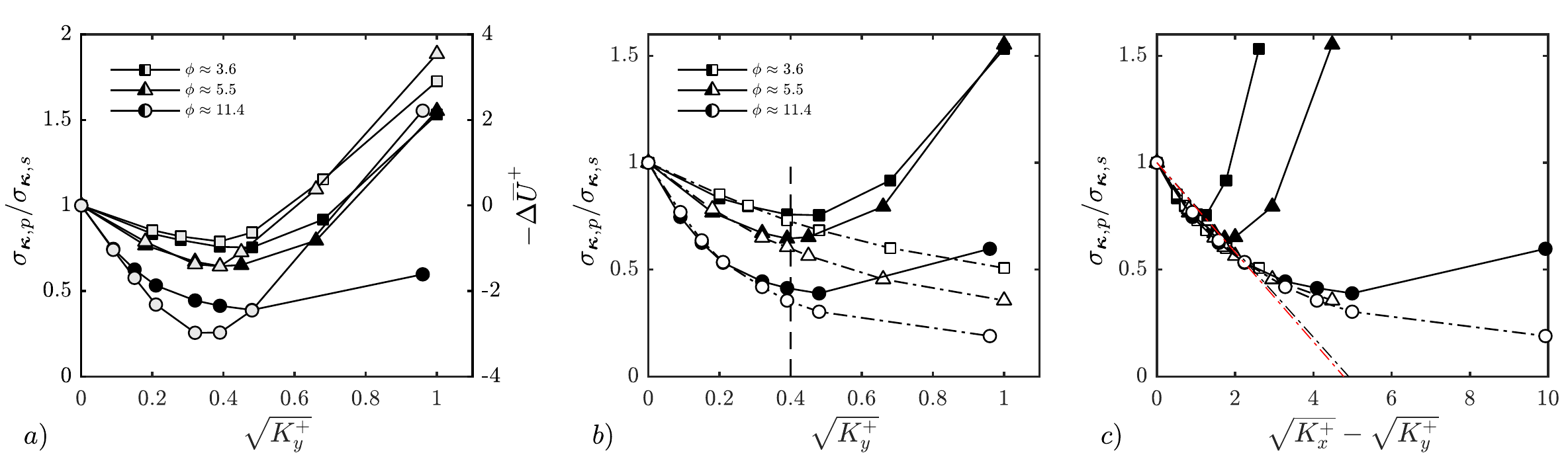}
	\caption{Predictions for the NW resolvent mode. (a) Comparison between normalized mode gain ($\sigma_{\bkv,p}/\sigma_{\bkv,s}$, black symbols) and drag reduction observed in DNS ($\Delta \mnU^+$, light gray symbols) by \citet{gomez-de-segura_garcia-mayoral_2019}, plotted as a function of $\protect\kpyy$.  The resolvent-based predictions shown in this panel are obtained using the velocity profiles from DNS. (b) Comparison between normalized gains predicted using the DNS mean profiles (filled symbols) and the synthetic mean profiles (open symbols) plotted as a function of $\protect\kpyy$. (c) Comparison between normalized gains predicted using the DNS mean profiles (filled symbols) and the synthetic mean profiles (open symbols) plotted as a function of $\protect\kpxx - \protect\kpyy$. The black and red dashed lines show linear fits to the initial decrease in normalized gain for the DNS- and synthetic-mean predictions, respectively. For all panels, the \protect\bsquare symbols represent substrates with $\phi_{xy} = 3.6$, \protect\btriangle symbols represent substrates with $\phi_{xy}= 5.5$, and \protect\bcircle symbols represent substrates with $\phi_{xy}= 11.4$.
	}\label{fig:nw_mode_gain}
\end{figure}

Previous studies have shown that the resolvent mode with streamwise wavelength $\llx = 10^3$, spanwise wavelength $\llz = 10^2$, and phase speed $c^+ = (\omega/\kxx)^+ = 10$ serves as a useful surrogate for the dynamically-important NW cycle characterized by the presence of quasi-streamwise vortices and alternating high- and low-speed streaks \citep{mckeon_sharma_2010,sharma2013coherent}.  The forcing-response gain (i.e., singular value) for this resolvent mode has also proven to be a useful predictor of control performance for both active \citep{luhar2014opposition,nakashima2017assessment,toedtli2019predicting} and passive \citep{luhar2015framework,luhar2016design,chavarin2020resolvent} techniques. Here, we evaluate whether it can serve as a useful reduced-complexity tool for the evaluation of anisotropic permeable substrates for passive drag reduction. Since the resolvent-based predictions are evaluated against DNS results from \citet{gomez-de-segura_garcia-mayoral_2019} carried out at $\Ret = 180$, the resolvent mode with streamwise wavenumber $\kxx = 2\pi \Ret/10^3 \approx 1.1$, spanwise wavenumber $\kzz = 2\pi \Ret/10^2 \approx 11$ and frequency $\omega = 10 \kxx \approx 11$ is used as a surrogate for the NW cycle.  

Figure~\ref{fig:nw_mode_gain}(a) shows the forcing-gain for the NW resolvent mode over the porous substrate $\sigma_{\bkv,p}$ normalized by the smooth-wall value $\sigma_{\bkv,s}$ at $\Ret = 180$ for all 22 substrates tested by \citet{gomez-de-segura_garcia-mayoral_2019}.  Resolvent-based gain predictions (filled black symbols) are compared against the drag reduction measured in DNS (filled gray symbols). Following \citet{gomez-de-segura_garcia-mayoral_2019}, the outward shift in the logarithmic region of the mean profile is used to quantify the change in drag: $\Delta \mnU^+ > 0$ denotes a decrease in drag and $\Delta \mnU^+ < 0$ denotes an increase in drag. For all three anisotropy ratios, there is qualitative agreement between NW mode suppression and drag reduction. For a given value of $\kpyy$, materials with higher anisotropy ratios are predicted to yield greater suppression of the NW resolvent mode, which is consistent with the drag reduction trends observed in DNS. Further, mode suppression and drag reduction both increase monotonically up to a value of $\kpyy \approx 0.4$. Agreement between drag reduction and mode suppression is particularly good for substrates with anisotropy ratios $\phi_{xy} = 3.6$ and $5.5$. For $\kpyy \gtrsim 0.4$, the normalized gain begins to increase again for the substrates with $\phi_{xy} = 3.6$ and $5.5$, which is consistent with DNS drag reduction trend. DNS results show an increase in drag relative to smooth-wall values for $\kpyy \gtrsim 0.55$ for the substrates with $\phi_{xy} = 3.6$ and $5.5$.  Resolvent-based predictions show that the gain of the NW mode over the permeable substrate exceeds the smooth-wall value beyond $\kpyy \gtrsim 0.7$ for these substrates.  For substrates with the higher anisotropy ratio, $\phi_{xy}= 11.4$, there are visible discrepancies between the predicted change in NW mode gain and the measured drag reduction, particularly for higher values of the wall normal permeability, $\kpyy$.  For instance, the increase in normalized gain beyond $\kpyy \gtrsim 0.4$ is less pronounced for the material with $\phi_{xy}= 11.4$.  For this substrate, DNS results show an increase in drag for $\kpyy \gtrsim 0.7$.  However, the resolvent predictions do not show an increase in gain beyond the smooth-wall value for the conditions tested (up to $\kpyy \approx 1$).

Note that the observed disagreement between NW mode gain and drag reduction for $\kpyy \gtrsim 0.4$ is not surprising.  As discussed in prior studies \citep{abderrahaman-elena_garcia-mayoral_2017,gg_sharma_gm_2018,gomez-de-segura_garcia-mayoral_2019}, the initial decrease in drag over anisotropic permeable substrates depends on the shift in the virtual origins for the streamwise mean flow and the turbulent cross-flow fluctuations produced by the quasi-streamwise vortices associated with the NW cycle.  As we show below, this effect is reproduced by the NW resolvent mode.  However, the increase in drag beyond $\kpyy \approx 0.4$ is associated with the emergence of spanwise rollers resembling KH vortices. This phenomenon cannot be captured by resolvent modes that serve as surrogates for the streamwise-elongated structures constituting the NW cycle.  We use the resolvent framework to test for the emergence of energetic spanwise rollers in Section~\ref{sec:KH_rollers}.

Figure~\ref{fig:nw_mode_gain}(b) shows normalized gain predictions made using the DNS mean profile (filled symbols) and the synthetic mean profile (open symbols) for the NW resolvent mode, plotted as a function of $\kpyy$. For values of $\kpyy \lesssim 0.4$, the normalized gain predictions are in close agreement.  For this range of wall-normal permeabilities, the normalized gain predictions made using the synthetic mean profile are within 6\% of those made using the DNS mean profile. However, for $\kpyy \gtrsim 0.4$, the synthetic profile predictions show a continued decrease in NW mode gain, while the DNS profile predictions show an increase in gain.  Discrepancy between the predictions for higher values of $\kpyy$ is consistent with the slip velocities shown in figure~\ref{fig:slip_figure}.  The synthetic profile agrees with the DNS mean profile for $\kpyy \lesssim 0.4$, but not beyond this value.  As discussed earlier, this is because higher wall-normal permeabilities give rise to spanwise rollers resembling KH vortices in the simulations.  The effect of the additional interfacial turbulence generated by these rollers is not captured in the eddy viscosity model (\ref{eqn:eddy_viscosity}).  Instead, the eddy viscosity model assumes smooth-wall like turbulence regardless of the value of $\kpyy$ and $\kpzz$.

Figure~\ref{fig:nw_mode_gain}(c) shows normalized gain predictions made using the DNS mean profile (filled symbols) and the synthetic mean profile (open symbols) for the NW resolvent mode, plotted as a function of $\kpxx - \kpyy$.  Since the permeable substrate model used here has identical wall-normal and spanwise permeabilities, the difference between streamwise and wall-normal permeability is also equal to the difference between streamwise and spanwise permeabilities, i.e., $\kpxx - \kpyy = \kpxx - \kpzz$.  Slip length arguments \citep{luchini_manzo_pozzi_1991,abderrahaman-elena_garcia-mayoral_2017} suggest that the initial decrease in drag over anisotropic permeable substrates is proportional to the difference between the streamwise and spanwise permeability length scales, which determine the virtual origins felt by the mean streamwise flow and turbulent cross-flow fluctuations, respectively.  In other words, we expect  $\Delta \mnU^+ \propto \kpxx-\kpzz$.  These arguments are supported by the simulation results of \citet{gomez-de-segura_garcia-mayoral_2019}, which show a linear decrease in drag that is proportional to $\kpxx-\kpzz$ for substrates with small permeabilities. The NW mode gains shown in figure~\ref{fig:nw_mode_gain}(c) agree well with this model. Specifically, the normalized NW-mode gains initially decrease linearly with $\kpxx - \kpyy  (= \kpxx - \kpzz)$ across all anisotropy ratios.  Moreover, there is little difference between the predictions made using the synthetic and DNS mean profiles. Linear relationships fitted to the initial decrease in normalized gain are nearly identical for the predictions based on the DNS profiles (dashed black lines) and those based on the synthetic mean profiles (dashed red lines).  Note that the predictions based on the synthetic mean profile all collapse together. This suggests that NW mode behavior is self-similar when appropriately normalized. Predictions made using the DNS profiles diverge from this self similarity when $\kpyy \gtrsim 0.4$. 

\begin{figure}
	\centering
	\includegraphics[width=0.95\textwidth]{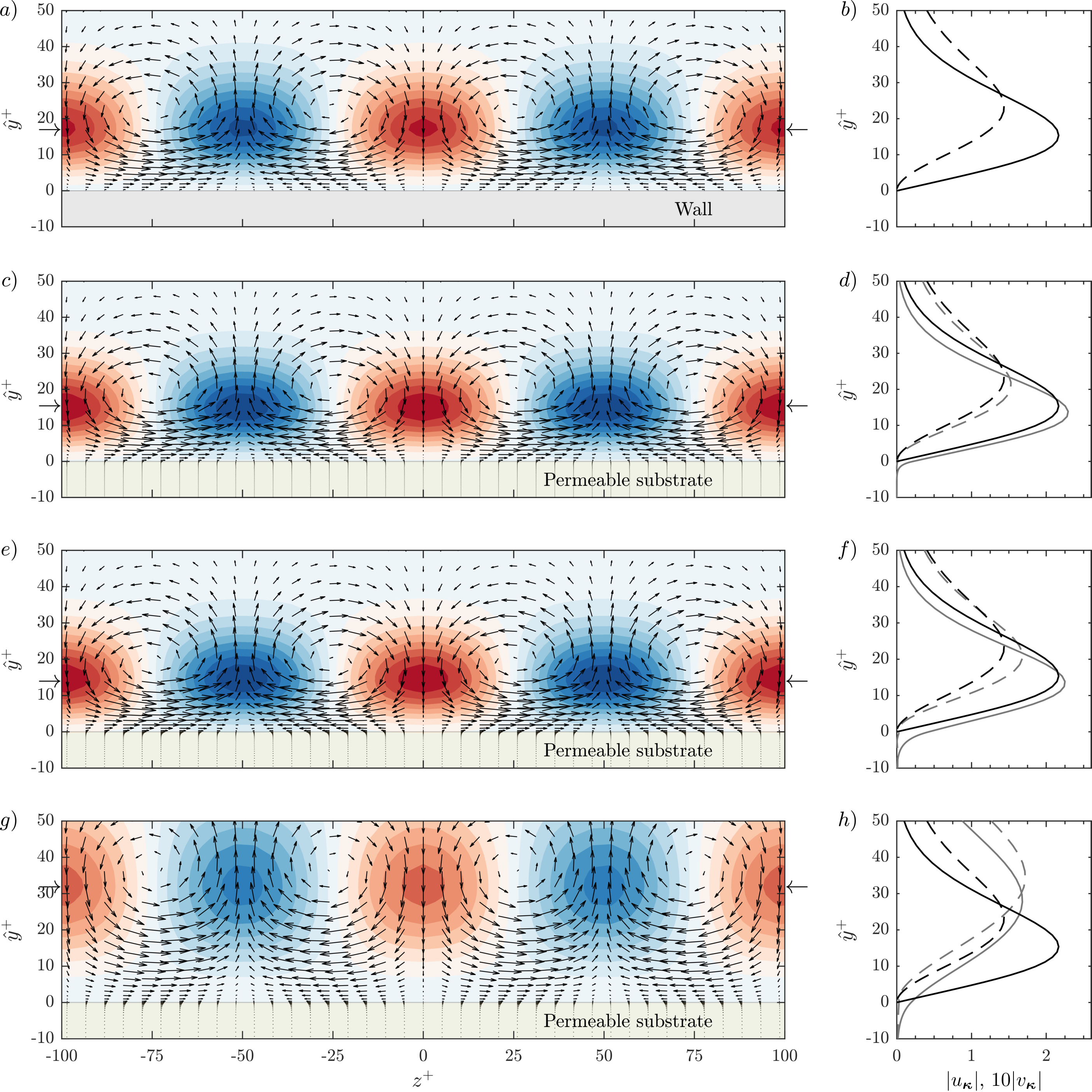}
	\caption{NW mode structure predictions made using the DNS velocity profiles for (a,b) the smooth-wall case, (c,d) the substrate with $\phi_{xy} = 5.5$ and $\protect\kpyy = 0.18$, and (e,f) the substrate with $\phi_{xy} = 5.5$ and $\protect\kpyy = 0.45$, and (g,h) the substrate with $\phi_{xy} = 5.5$ and $\protect\kpyy = 1.0$.  In (a,c,e,f), the shading shows normalized contours of positive (red) and negative (blue) streamwise velocity normalized by the maximum value. The vectors show the wall-normal and spanwise velocity fluctuations.  In (b,d,f,g), the solid lines show the magnitude of the streamwise velocity for this resolvent mode, $|u_{\bkv}|$ and the dashed lines show the wall-normal velocity magnitude multiplied by a factor of ten, $10 |v_{\bkv}|$.  The black lines represent the smooth-wall case while the gray lines represent the permeable substrates.  These plots make use of the shifted coordinate  $\hat{y}=1+y$, such that $\hat{y}=0$ represents the location of the smooth wall or the porous interface.}\label{fig:mode_shift}
\end{figure} 

%

Figure~\ref{fig:mode_shift} provides physical insight into the observed changes in mode gain. The structure of the NW resolvent mode for the smooth-wall case is shown in figure~\ref{fig:mode_shift}(a).  By construction, the mode structure shows alternating regions of positive and negative velocity fluctuations with spanwise wavelength $\llz = 10^2$. The wall-normal and spanwise velocity fluctuations show the presence of counter-rotating streamwise vortices. Regions of downwelling ($v^\prime$ towards wall) coincide with regions of high streamwise velocity (red shading) and regions of upwelling ($v^\prime$ away from wall) coincide with regions of low streamwise velocity (blue shading). These observations are consistent with known features of the NW cycle \citep{robinson1991coherent,waleffe1997self,jimenez_pinelli_1999}.  When the solid wall is replaced by anisotropic permeable substrates with $\phi_{xy} = 5.5$, there are some subtle but important changes in mode structure.  For the permeable substrates that yield mode suppression---and drag reduction---there is a downward shift in the velocity fluctuations and the mode is \textit{compressed} in the wall-normal direction (see Figs.~\ref{fig:mode_shift}(c,d) for $\kpyy = 0.18$ and Figs.~\ref{fig:mode_shift}(e,f) for $\kpyy = 0.45$).  To explain why these changes lead to mode suppression, we briefly review two mechanisms responsible for high gain in the resolvent framework \citep{mckeon_sharma_2010,mckeon2017engine}.  First, resolvent modes tend to localize around the critical layer, $y_c$, where the mode speed matches the local mean velocity, $\mnU(y_c) = c = \omega/\kxx$.  This minimizes the term $-i\omega + i\kxx \mnU$ inside the inverted operator shown on the right-hand side of (\ref{eqn:mod_resolvent}).  Second, energy is extracted from the mean flow via the interaction between the wall-normal velocity fluctuations and the mean shear, i.e., the so-called \textit{lift-up} mechanism that leads to energy transfer from the mean flow to the fluctuations via the term $-\overbar{u^\prime v^\prime} (d\mnU/dy)$ in the turbulent kinetic energy equation.  As the streamwise permeability of the substrate increases, there is a shift in the virtual origin of the mean profile and so the critical layer location moves closer to the permeable interface, from $\hat{y}_c^+ \approx 13$ for the smooth-wall case to $\hat{y}_c^+ \approx 9.7$ for the case with $\kpyy = 0.45$.  Here, $\hat{y} = 1+y$ such that $\hat{y}=0$ represents the porous interface at $y=-1$ (see figure~\ref{fig:channel_dim}).  As noted in the introduction, this shift in the mean profile depends on $\kpxx$.  A downward shift in the critical layer location causes the NW resolvent mode to localize closer to the porous interface (see $\rightarrow-\leftarrow$ in figure~\ref{fig:mode_shift}).  However, the downward shift for the NW mode is constrained by the low wall-normal and spanwise permeabilities, $\kpyy = \kpzz$, which leads to the observed wall-normal compression in mode structure. 
This wall-normal compression in mode structure reduces the energy extracted from the mean flow, which depends on the wall-normal integral of the product of the Reynolds shear stress and mean shear, i.e.,  $\propto \int \Real(-u_{\bkv}^* v_{\bkv}) (d\mnU/dy) dy$.  Note that $\Real(\cdot)$ represents the real component and so the Reynolds shear stress contribution from the NW resolvent mode is proportional to $\Real(-u_{\bkv}^* v_{\bkv})$ \citep{luhar2014opposition}.  A reduction in energy extraction from the mean flow leads to a reduction in mode gain.  Importantly, this observation is consistent with the interpretation that drag reduction is only expected if the slip length for the mean streamwise flow $\ell_x^+ \propto \kpxx$ is larger than that for the turbulent cross-flows, $\ell_z^+ \propto \kpzz$, i.e., if the virtual origin for the mean flow is offset from the virtual origin for the turbulence. Additional resolvent-based predictions for $\kpzz > \kpxx$ ($\ell_z^+ > \ell_x^+$) lead to an increase in mode gain (data not shown).

Figures~\ref{fig:mode_shift}(g,h) show predictions for NW mode structure for the anisotropic permeable substrate with $\phi_{xy} = 5.5$ and $\kpyy = 1.0$ that leads to an increase in drag and forcing-response gain (see figure~\ref{fig:nw_mode_gain}).  Compared to the cases that lead to mode suppression and drag reduction, the wall-normal footprint for the NW mode increases substantially for this case.  This can be attributed to the change in the DNS mean profiles for $\kpyy \gtrsim 0.4$, for which spanwise-coherent structures resembling KH vortices emerge over the permeable substrate.  The emergence of these rollers leads to a substantial increase in turbulent mixing which causes the mean profile to flatten in the interfacial region \citep{gomez-de-segura_garcia-mayoral_2019}.  For the substrate with $\phi_{xy} = 5.5$ and $\kpyy = 1.0$, this pushes the critical layer for the NW mode outwards to $\hat{y}_c^+ \approx 26$. This means that the NW mode is again able to localize around the critical layer, and there is no vertical compression in mode structure.  This allows for greater energy extraction from the mean flow and leads to the observed increase in NW mode gain over the material with $\phi_{xy}= 5.5$ and $\kpyy = 1.0$  in figure~\ref{fig:nw_mode_gain}(a). 

Note that the NW mode gain decreases monotonically with increasing $\kpxx - \kpyy$ for predictions made using the synthetic mean profiles (see figure~\ref{fig:nw_mode_gain}(c)). This is because the synthetic mean profiles assume no change in turbulence characteristics near the interface relative to smooth-wall conditions, i.e., they do not account for the emergence of spanwise rollers and turbulence penetration into the porous medium.  As shown in figure~\ref{fig:nw_mode_gain}(a), NW mode gain begins to increase again for $\kpyy \gtrsim 0.4$ when the DNS mean profiles are used in the resolvent operator.  In the DNS, spanwise rollers resembling KH vortices emerge beyond this value of wall-normal permeability and alter the form of the mean profile. 

The preceding discussion assumes that the phase speed of the near-wall structures ($c^+ \approx 10$) remains unchanged relative to smooth wall conditions.  It could also be argued that the phase speed changes over the permeable medium to reflect the interfacial slip velocity, which depends on the streamwise slip length ($\mnU_s^+ \approx \ell_x^+$ for small $\ell_x^+$), and turbulence penetration into the permeable medium, which depends on the spanwise slip length ($\ell_z^+$). To account for these effects, the phase speed can be modified to $c^+ = 10 + \ell_x^+ - \ell_z^+ \approx 10 + \kpxx - \kpzz$. Figure~\ref{fig:nw_mode2_gain} in Appendix~\ref{appA} shows additional predictions for NW mode gain made using this modified phase speed. These additional predictions are broadly similar to the predictions shown in figure~\ref{fig:nw_mode_gain} for $c^+ = 10$. Specific observations are discussed further in Appendix~\ref{appA}.

Together, the predictions shown in figure~\ref{fig:nw_mode_gain} and figure~\ref{fig:mode_shift} confirm that suppression of the dynamically-important NW cycle requires high streamwise permeability and low spanwise/wall-normal permeabilities.  In other words, materials with high anisotropy ratios ($\phi_{xy} \gg 1$) are expected to yield the largest suppression of NW turbulence.  However, the absolute value of the wall-normal permeability also plays an important role in dictating drag reduction performance.  Specifically, the emergence of energetic spanwise coherent rollers is dictated primarily by $\kpyy$. 

\subsection{Spanwise-Coherent Resolvent Modes}\label{sec:KH_rollers}

\begin{figure}
	\centering
	\includegraphics[width = \textwidth]{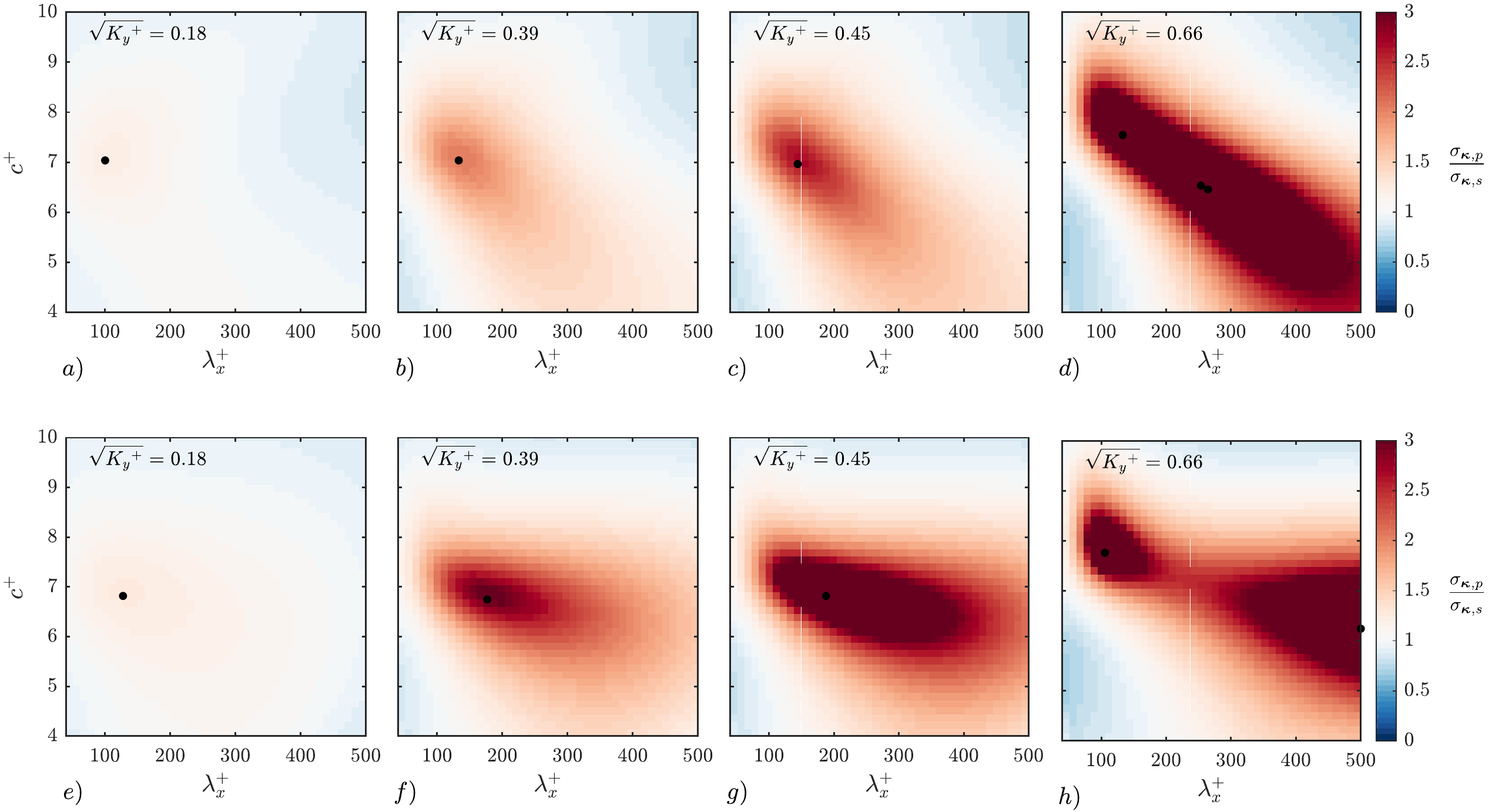}
	\caption{Normalized gain for spanwise-coherent structures plotted as a function of streamwise length $\llx$ and wave speed $c^+$ for substrates with an anisotropy $\phi_{xy}= 5.5$.  These predictions make use of the synthetic mean profile. Panels (a)-(d) represent structures with spanwise wavenumber $\kzz\approx 2.3$ ($\protect\llz=500$).  Panels (e)-(h) represents structures with spanwise wavenumber $\kzz=0$. Red and blue shading  respectively represent mode amplification and suppression relative to the smooth-wall case. Wall-normal permeability increases from left to right. 
	Structures with the largest amplification are labeled with a ($\bullet$) marker.}\label{fig:roller_gain_55case}	
\end{figure}


In this section, we use the resolvent analysis framework to evaluate the conditions in which energetic spanwise-coherent rollers resembling Kelvin-Helmholtz vortices emerge over anisotropic permeable substrates. Snapshots of the flow field from the simulations of \citet{gomez-de-segura_garcia-mayoral_2019} show that for $\kpyy < 0.4$, the flow is dominated by streamwise-elongated structures associated with the NW cycle. As the wall-normal permeability increases to $\kpyy \gtrsim 0.4$ the flow becomes dominated by spanwise coherent rollers with an approximate streamwise spacing of $\lambda_x^+ \approx 100-300$. Further, using momentum balance arguments, \citet{gomez-de-segura_garcia-mayoral_2019} showed that the Reynolds stress contribution from structures with $\lambda_x^+ \approx 70 - 320$ and $\lambda_z^+ \geq 120$ is responsible for the increase in drag observed for substrates with $\kpyy \gtrsim 0.4$. We note that spanwise-coherent rollers have also been observed in previous simulations over isotropic porous materials \citep{breugem2006influence,rosti2015direct}.

Linear stability analyses successfully predict the emergence of Kelvin-Helmholtz vortices in channel flows over anisotropic permeable substrates \citep{abderrahaman-elena_garcia-mayoral_2017,gg_sharma_gm_2018}. \citet{gg_sharma_gm_2018} also show that the wall-normal permeability becomes the driving parameter for the onset of KH rollers for streamwise-preferential substrates of large depth, i.e., substrates with $\kpxx \gg \kpyy$ and $h^+ \gg \kpyy$. However, models based on linear stability theory do not accurately predict the threshold value of $\kpyy$ for the onset of the rollers or the wavelength of the rollers.  Linear stability analyses predict structures with streamwise wavelength $\llx \approx 50-70$ to be most unstable, while DNS results suggest that structures with $\llx \approx 150$ are most energetic initially.  

Figure~\ref{fig:roller_gain_55case} shows resolvent-based predictions for the normalized amplification of spanwise-coherent structures over the anisotropic permeable substrate with $\phi_{xy}= 5.5$ and varying $\kpyy$.  Based on the DNS results of \citet{gomez-de-segura_garcia-mayoral_2019}, we limit ourselves to the evaluation of structures with streamwise wavelength $\llx\in[50,500]$ and mode speed $c^+\in[4,10]$. In addition, we consider structures that are infinitely long in the spanwise direction ($\kzz = 0$) as well as structures with a finite, but large, spanwise wavelength ($\kzz \approx 2.3$; $\llz = 500$). 

Figures~\ref{fig:roller_gain_55case}(a,e) show that for low wall-normal permeability, $\kpyy=0.18$, resolvent analysis predicts minimal amplification of spanwise-coherent structures over the anisotropic permeable substrate. The maximum increase in gain relative to smooth-wall values is less than $25\%$, i.e., $\sigma_{\bkv,p}/\sigma_{\bkv,s}<1.25$ for both $\kzz = 0$ and $\kzz \approx 2.3$.  Moreover, the gain for these spanwise-coherent structures is also small in absolute terms. This is consistent with the simulation results of \citet{gomez-de-segura_garcia-mayoral_2019}, which show that the Reynolds stress contributions from structures with $\llx\approx 70-300$ and $\llz\gtrsim120$ are negligible for materials with $\kpyy\lesssim 0.31$.  In other words, spanwise-coherent modes do not have a significant effect on drag for these conditions.  

However, the amplification of these spanwise-coherent structures increases significantly for $\kpyy \ge 0.39$. Figures~\ref{fig:roller_gain_55case}(b,f) show a clear increase in the size and intensity of the region showing mode amplification. For $\llz=500$, the region of highly-amplified structures is localized around $(\llx,c^+)\approx(130,7)$ and the normalized gain for the most amplified structure is $\sigma_{\bkv,p}/\sigma_{\bkv,s} \approx 2$. For spanwise constant structures, the region of high amplification is localized around $(\llx,c^+)\approx(180,6.7)$ and extends to structures with $\llx \approx 300$.  The normalized gain for structures in this region is as high as $\sigma_{\bkv,p}/\sigma_{\bkv,s} \approx 3$, indicating a 200\% increase in gain relative to smooth-wall values.  The extent and intensity of these high-gain regions continue to increase as wall-normal permeability increases further, as shown in Figs.~\ref{fig:roller_gain_55case}(c,d) and (g,h).  Indeed, for the case with $\kpyy = 0.66$ shown in figure~\ref{fig:roller_gain_55case}(d), the region of high amplification extends from $\llx \approx 100-500$ and $c^+ \approx 4-9$.  The highest-amplification structures in this region show normalized gains $\sigma_{\bkv,p}/\sigma_{\bkv,s} \ge 100$.  

The substantial increase in amplification of spanwise-coherent structures for $\kpyy \ge 0.39$ is again consistent with the DNS results of \citet{gomez-de-segura_garcia-mayoral_2019}, which show that the drag-reducing performance of anisotropic substrates begins to degrade for $\kpyy \ge 0.39$ due to the additional Reynolds shear stresses contributed by spanwise coherent structures with $\llx \approx 70-300$ and $\llz > 120$.  The DNS results also show the emergence of a new peak in the premultiplied spectra for wall-normal velocity near $\llx \approx 150$ for $\kpyy = 0.39$.  This is very close to the streamwise wavelength of the most amplified modes in Figs.~\ref{fig:roller_gain_55case}(b,f).  Moreover, the streamwise extent of the high-intensity region in the DNS premultiplied spectra increases with increasing $\kpyy$, which is similar to the expansion of the high-amplification region evident in Figs.~\ref{fig:roller_gain_55case}(c,d,g,h).

\begin{figure}
	\centering
	\includegraphics[width=\textwidth]{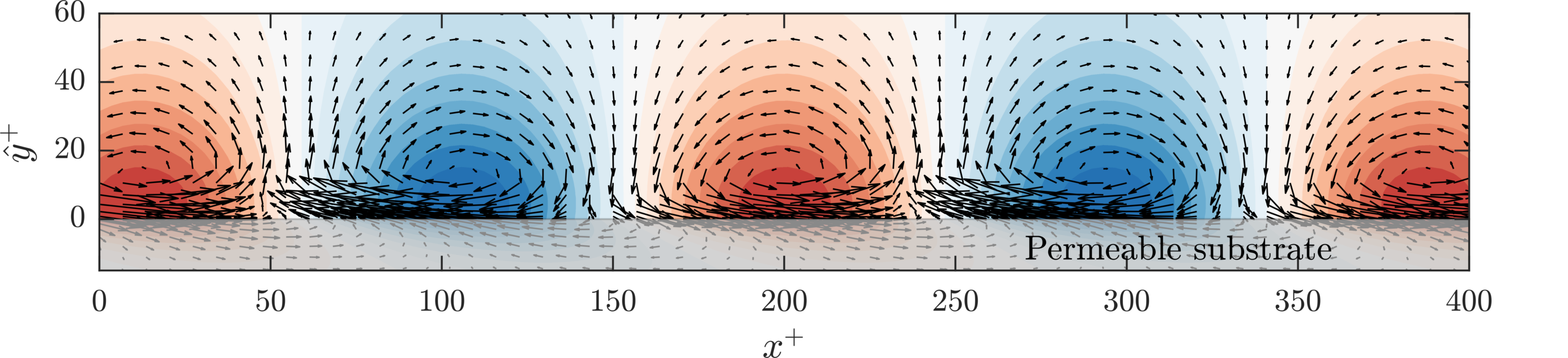}
	\caption{Flow structure associated with the most amplified spanwise-constant resolvent mode over the permeable substrate with $\phi_{xy}=5.5$ and $\protect\kpyy=0.39$; see $\protect\bullet$ in figure~\ref{fig:roller_gain_55case}(f).  The red and blue shaded contours show regions of positive and negative pressure, normalized by the maximum value.}\label{fig:kh_mode_shape_inf}
\end{figure}

Figure~\ref{fig:kh_mode_shape_inf} shows a snapshot of the flow field associated with the highest-gain resolvent mode in figure~\ref{fig:roller_gain_55case}(f) with $\llx \approx 180$ and $c^+ \approx 6.7$.  As expected for Kelvin-Helmholtz type vortices, the structure shows the presence of counter-rotating rollers in the $x-y$ plane, localized near the porous interface.  Regions of prograde rotation (i.e., in the direction of the mean shear) are associated with negative pressure fluctuations and regions of retrograde rotation are associated with positive pressure fluctuations.  Moreover, the velocity and pressure fields associated with these rollers span a significant portion of the buffer region of the flow, which is consistent with DNS observations.

\begin{figure}
	\centering
	\includegraphics[width=0.9\textwidth]{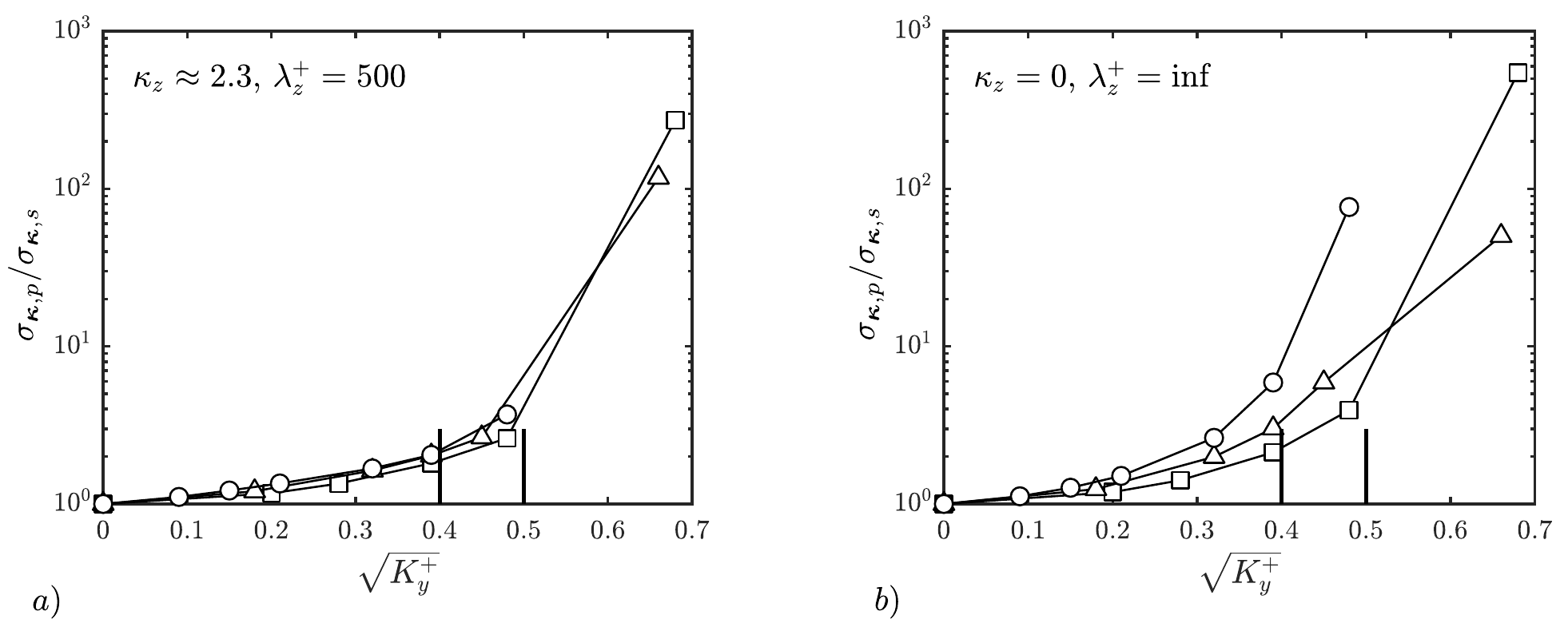}
	\caption{Comparison of resolvent-based gain predictions for spanwise-coherent structures with anisotropy ratios $\phi_{xy}=3.6$ (\,\,\protect\bsquare),  $5.5$ (\,\,\protect\btriangle), and $11.4$ (\,\,\protect\bcircle). The maximum normalized gain obtained for resolvent modes with $\protect\llx\in[50,500]$ and $c^+ \in [4,10]$ is shown as a function of wall-normal permeability for (a) $\kzz \approx 2.3$ ($\llz = 500$) and for (b) $\kzz = 0$ ($\llz = \infty$). All of these predictions were obtained using the synthetic mean profiles.}\label{fig:kh_all_500_inf}
\end{figure}

Importantly, the model predictions shown in figure~\ref{fig:roller_gain_55case} for $\phi_{xy} = 5.5$ are representative of what is observed over the permeable substrates with $\phi_{xy} = 3.6$ and $\phi_{xy} = 11.4$ as well.  This is illustrated well in figure~\ref{fig:kh_all_500_inf}, which shows the normalized gain for the \textit{most amplified} resolvent mode in the window spanning $\llz\in[50,500]$ and $c^+\in[4,10]$, for both $\kzz \approx 2.3$ and $\kzz = 0$.  For all three anisotropy ratios, the normalized gain show a sharp increase for $\kpyy \gtrsim 0.4$.  For structures with $\llz = 500$ ($\kzz \approx 2.3$), model predictions for all three anisotropy ratios collapse together nicely (figure~\ref{fig:kh_all_500_inf}(a)).  For spanwise-constant structures ($\kzz = 0$), there is a bit more scatter in the normalized gain values but the overall trend remains similar.  

The DNS results in figure~\ref{fig:nw_mode_gain}(a) clearly show drag reduction performance deteriorating for $\kpyy \gtrsim 0.4$ for all three anisotropy ratios. The flow field snapshots, velocity spectra, and momentum-based analyses pursed by \citet{gomez-de-segura_garcia-mayoral_2019} attribute this deterioration in performance to the emergence of high-gain spanwise-coherent structures resembling Kelvin-Helmholtz vortices. The results presented in this section confirm that resolvent analysis---based on synthetic mean profiles---is able to successfully predict the emergence of such spanwise-coherent structures over anisotropic permeable substrates. Since the synthetic mean profiles are generated used an eddy viscosity model, these results also show that the resolvent framework can generate \textit{a priori} predictions for whether a porous substrate with known $\bfK$ will give rise to KH rollers.  Moreover, the resolvent-based predictions are in better quantitative agreement with the DNS results than linear stability predictions.  Specifically, resolvent analysis yields better predictions for the wall-normal permeability threshold beyond which such structures become important ($\kpyy \approx 0.4$) as well as the streamwise wavelength of the structures that emerge first ($\llx \approx 150$).

\section{Conclusion}\label{sec:conclusion}
Recent theoretical efforts and numerical simulations show that anisotropic permeable substrates have the potential to yield significant drag reduction in wall-bounded turbulent flows. Specifically, numerical simulations by \citet{gomez-de-segura_garcia-mayoral_2019} indicate that a drag reduction of up to 25\% can be achieved by appropriately tuning the streamwise, wall-normal, and spanwise permeabilities of such substrates. The mechanism through which these substrates reduce drag is similar to that for riblets, and can be rationalized in terms of the offset in the virtual origin felt by the mean flow, which depends on streamwise permeability, and the virtual origin felt by the turbulent fluctuations, which depends primarily on spanwise permeability. Maximum drag reduction is limited by the appearance of spanwise rollers resembling Kelvin-Helmholtz vortices.  The emergence of these vortices is linked to a relaxation in the wall-normal permeability.  The extended resolvent analysis framework described in this paper reproduces all of these features with minimal computation.  

As shown in \S\ref{sec:nw_resolvent_mode}, the gain for a single resolvent mode that serves as a surrogate for the near-wall cycle reproduces the initial drag reduction trends observed in numerical simulations.  Model predictions show that the reduction in gain for this mode depends on the difference between the streamwise and spanwise permeability length scales, $\kpxx - \kpzz$, such that substrates with higher anisotropy ratios lead to greater mode suppression. This is consistent with the trends observed in numerical simulations.  Moreover, the link between resolvent mode gain and drag reduction also provides a complementary explanation to the slip-length based arguments used in previous studies. Specifically, resolvent-based predictions suggest that the difference between $\kpxx$ and $\kpzz$ leads to a wall-normal compression of the quasi-streamwise vortices associated with the near-wall cycle, and this compression limits energy extraction from the mean flow.  Further, as we show in \S\ref{sec:KH_rollers}, resolvent analysis also predicts the emergence of energetic spanwise rollers with streamwise wavelength $\lambda_x^+ \approx 150$ as the wall-normal permeability increases beyond $\kpyy \approx 0.4$.  These features are quantitatively consistent with simulation results. 

One weakness of the extended resolvent framework developed here is the requirement of a mean velocity profile over the permeable substrate.  For turbulent flows over smooth walls, mean profile estimates can be generated using a variety of simplified models or obtained from previous simulations and experiments.  However, such models or data are not readily available for flows over anisotropic permeable substrates.  The form of the mean profile can have a significant effect on resolvent-based predictions. Indeed, the comparison between predictions made using the mean profiles obtained in DNS and those generated using an eddy viscosity model presented in \S\ref{sec:nw_resolvent_mode} shows important differences in the gain for the near-wall mode for $\kpyy \gtrsim 0.4$, i.e., when drag reduction performance begins to deteriorate.  This is because the emergence of the spanwise rollers leads to a substantial increase in interfacial turbulence. This effect is not captured by the eddy viscosity model used to generate the synthetic profiles, which assumes that the turbulence remains smooth-wall like and does not penetrate the permeable substrate (see \S\ref{sec:mean_vel_methods}). Yet, resolvent analysis with the synthetic mean profile does show the emergence of these spanwise rollers beyond $\kpyy \approx 0.4$.  In other words, resolvent analysis can generate useful \textit{a priori} predictions even with a synthetic mean profile. Near-wall mode gain can be used as a measure of initial drag reduction performance, and the emergence of spanwise-coherent rollers can be used to estimate the point at which performance is likely to deteriorate. Since evaluation of these features does not require significant computation, resolvent analysis can serve as a useful design tool for more complex permeable substrates.  For instance, resolvent analysis could be used to pursue formal optimization efforts for the full permeability tensor.  

We also recognize that the volume-averaged Navier-Stokes equations used in this study involved significant simplifying assumptions (see \S\ref{sec:gov_mods}). These simplifying assumptions were made primarily to allow for a direct comparison with previous simulation results.  Moving forward, resolvent analysis could also be used to consider how the inclusion of inertial effects (e.g., via the Forchheimer term) or more complex interfacial boundary conditions (e.g., involving stress jumps, or more complex slip and transpiration models) is likely to affect control performance.


	
\begin{acknowledgments}
This material is based on work supported by the Air Force Office of Scientific Research under awards FA9550-17-1-0142 (Program Manager: Dr. Gregg Abate) and FA9550-19-1-7027 (Program Manager: Dr. Douglas Smith).  Declaration of interests: the authors report no conflict of interest.
\end{acknowledgments}

\appendix\section{}\label{appA}

\begin{figure}
	\includegraphics[width=\textwidth]{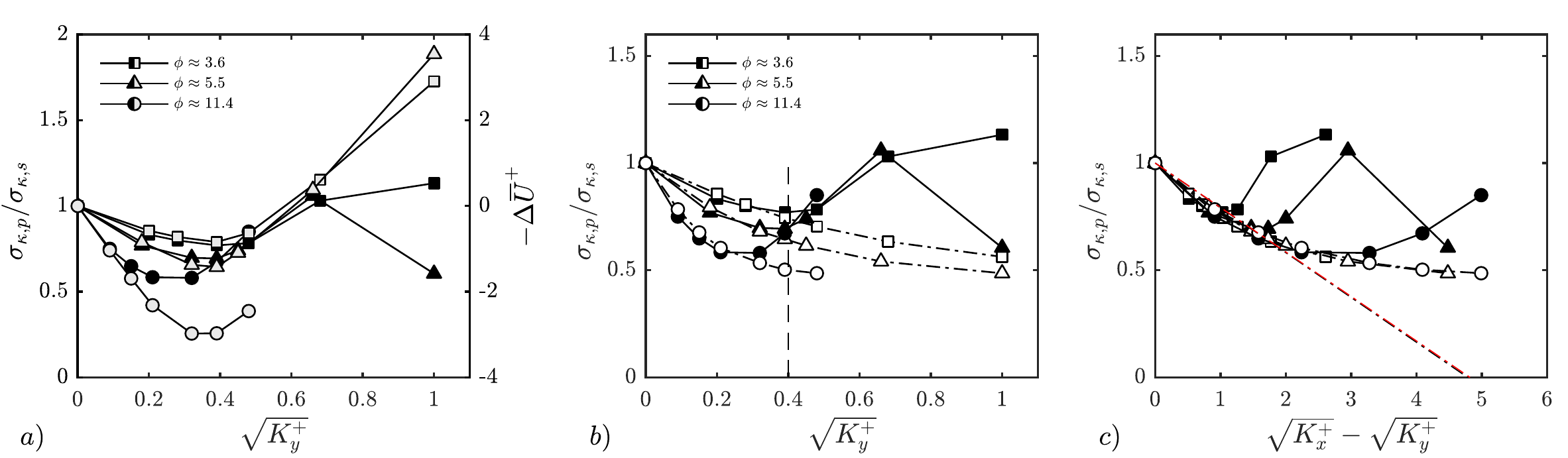}
	\caption{Predictions for the NW resolvent mode with a modified phase speed of $c^+ \approx 10 + \protect\kpxx - \protect\kpzz$. (a) Comparison between normalized mode gain ($\sigma_{\bkv,p}/\sigma_{\bkv,s}$, black symbols) and drag reduction observed in DNS ($\Delta \mnU^+$, light gray symbols) by \citet{gomez-de-segura_garcia-mayoral_2019}, plotted as a function of $\protect\kpyy$.  The resolvent-based predictions shown in this panel are obtained using the velocity profiles from DNS. (b) Comparison between normalized gains predicted using the DNS mean profiles (filled symbols) and the synthetic mean profiles (open symbols) plotted as a function of $\protect\kpyy$. (c) Comparison between normalized gains predicted using the DNS mean profiles (filled symbols) and the synthetic mean profiles (open symbols) plotted as a function of $\protect\kpxx - \protect\kpyy$. The black and red dashed lines show linear fits to the initial decrease in normalized gain for the DNS- and synthetic-mean predictions, respectively. For all panels, the \protect\bsquare symbols represent substrates with $\phi_{xy} = 3.6$, \protect\btriangle symbols represent substrates with $\phi_{xy}= 5.5$, and \protect\bcircle symbols represent substrates with $\phi_{xy}= 11.4$.}
	\label{fig:nw_mode2_gain}
\end{figure}

As noted earlier, the predictions shown in \S\ref{sec:nw_resolvent_mode} assume that the phase speed of the resolvent mode that serves as a surrogate for the NW cycle ($c^+ \approx 10$) remains unchanged over the permeable substrate.  Figure~\ref{fig:nw_mode2_gain} shows additional predictions for mode gain assuming the phase speed change with substrate permeability as $c^+ = 10 + \ell_x^+ - \ell_z^+ \approx 10 + \kpxx - \kpzz$.  In effect, this modification accounts for the interfacial slip velocity, which depends on the streamwise slip length $\ell_x^+ \approx \kpxx$, and the virtual origin for the turbulence inside the permeable substrate, which depends on the spanwise slip length $\ell_z^+ \approx \kpzz$. 

The modified phase speed does not lead to significant changes in mode gain compared to the predictions shown in figure~\ref{fig:nw_mode_gain} for $c^+ \approx 10$. Figure~\ref{fig:nw_mode2_gain}(a) shows that mode gain predictions made using the DNS mean profile are consistent with drag reduction trends obtained in DNS. Compared to the predictions shown in figure~\ref{fig:nw_mode_gain}(a), singular value ratios for the substrates with $\phi_{xy}=3.6$ and $\phi_{xy}=5.5$ show slightly better agreement with the drag reduction trends from DNS for $\kpyy < 0.6$.  However, for $\kpyy \gtrsim 0.6$, the predictions are more scattered with the modified phase speed. For the substrate with $\phi_{xy}= 11.4$, the modified phase speed leads to reduced mode suppression.  Further, the deterioration in performance---in terms of mode suppression---commences at a lower value of wall-normal permeability, $\kpyy \approx 0.2$, compared to the threshold observed in DNS. 

Figure~\ref{fig:nw_mode2_gain}(b) shows that the predictions made using the synthetic mean profile are consistent with those made using the DNS mean profiles for small values of $\kpyy$, i.e., until the emergence of spanwise rollers resembling Kelvin-Helmholtz vortices in the DNS. Figure~\ref{fig:nw_mode2_gain}(c) shows that the initial reduction in mode gain depends primarily on $\kpxx-\kpzz$, even with the modified phase speed. These observations are again consistent with the predictions shown in figure~\ref{fig:nw_mode_gain}(b,c). However, linear fits to the initial decrease in mode gain show that the modified phase speed leads to less pronounced mode suppression; see red and black dashed lines in figures~\ref{fig:nw_mode_gain}(c) and \ref{fig:nw_mode2_gain}(c).

\bibliographystyle{jfm}\bibliography{ac_ml_2019}%
\end{document}